\documentclass[11pt]{article}

\usepackage[margin=1.1in]{geometry}
\usepackage[utf8]{inputenc}
\usepackage[T1]{fontenc}
\usepackage{lmodern}
\usepackage{microtype}
\usepackage{amsmath,amssymb}
\usepackage{booktabs}
\usepackage{graphicx}
\usepackage[numbers,sort&compress]{natbib}
\PassOptionsToPackage{hyphens}{url}
\usepackage[hidelinks]{hyperref}
\hypersetup{pdftitle={The Civilization Framework: Sovereign-Anchored Communication Between Personal Multi-Agent Systems},
  pdfauthor={Guangjun Liu}}
\usepackage{enumitem}
\setlist{topsep=2pt,itemsep=1pt}

\newcommand{\framework}{Civilization Framework}
\newcommand{\embassy}{Embassy Protocol}

\title{The Civilization Framework: Sovereign-Anchored\\
Communication Between Personal Multi-Agent Systems}

\author{Guangjun Liu\\
\small New York University\\
\small \texttt{gl2980@nyu.edu}}

\date{Preprint, September 2026}

\begin{document}
\maketitle

\begin{abstract}
In professional collaboration, humans have quietly become the transport layer
between AI systems: one person's assistant produces an artifact, the person
relays it through chat or email, and the counterpart re-feeds it to their own
assistant, losing environment details, versions, and constraints at every hop.
We argue that the natural unit of AI-mediated collaboration is not the
individual agent but what we call a \emph{civilization}: one human sovereign
together with their persistent ledger and the interchangeable agents that
act on it. The name is deliberate on two grounds: in the AI era a
single person commands civilization-scale knowledge and, with agents,
civilization-scale capacity to act; and the unit has a civilization's
defining structure, outliving every agent that serves it while
preserving memory and norms.
We present the \framework{}, a conceptual architecture in which
civilizations, not agents, are the addressable parties, and the
\embassy{}, a carrier-agnostic overlay in which a resident ledger endpoint
receives communication asynchronously (store-and-forward), any online agent
of the receiving civilization can claim and process it, and commitment
state recorded on both ledgers, not message delivery, constitutes the ground
truth of the exchange. Authority in the framework derives from memory:
an agent's power to represent its civilization is capped by the scope of
memory it can access, externalized to counterparties through signed
credentials, and separated from a civilization-level reputation earned
across exchanges. We further identify a hazard acute in AI-to-AI
communication, the \emph{temporal-weight} effect, in which information that
arrives first acquires unearned authority, and test it in one frontier
model in a preregistered experiment of 1{,}908 trials. The observed
pattern is large: with verification capability removed, an incorrect
upstream claim arriving first captures 54.2\% of answers (4.2\% under
full verification), while the same claim arriving after the receiver
has sealed its own answer captures 31.6\%, separating arrival order
from mere exposure (the first-arrival and sealed-answer prompt shells
are not length-matched, so part of that difference may reflect shell
form; see \S\ref{sec:evaluation}), and both registered question-set
specifications agree on these two verdicts (the exclusion
specification is registered in advance as under-powered). Two
secondary results, the mitigation
from instruction-level provenance labeling and the equivalence of
sealed-answer accuracy, are specification-dependent and hold only under
the all-questions specification. Because the registered objective
read-back of tool use failed its call-budget condition, the
registration classifies this round as inconclusive and every result
above, primary and secondary alike, is reported as exploratory; a
replication with harness-enforced budgets is planned. The pattern
suggests protocol-level mitigations at the capability and procedure
level; that reading is an interpretation, not a registered result. The framework is grounded in a working reference implementation
of its intra-civilization layer and
positioned against forty years of agent-communication research, from
speech-act semantics through social commitments to today's A2A and MCP
protocol stack.
\end{abstract}

\section{Introduction}
\label{sec:intro}

Consider two engineers collaborating on a deployment. Each works with a
capable AI assistant. The backend engineer's assistant produces an
installation guide; the engineer pastes it into a chat application; the
frontend engineer forwards it to her own assistant, which discovers that
the guide omits the local environment configuration and the pinned
dependency versions. A question travels back through both humans. Three
round trips later, the missing constraint is finally identified. Every hop
crossed two human relays, each asynchronous, forgetful, and lossy.

We call this the \emph{porter problem}: in AI-assisted professional work,
humans have become the transport layer between AI systems. The artifacts
humans relay (documents, chat messages) are lossy snapshots curated for
human reading; the information that AI systems actually need to
synchronize (environment fingerprints, version manifests, implicit
constraints) is precisely what such snapshots omit. Meanwhile the AI
systems on both ends maintain rich, structured state about their
respective sides, state that direct synchronization could reconcile
without the human round trips.

The cost has begun to be named and measured outside academia: an industry
survey describes employees as ``human middleware'' who lose the better
part of a workday each week relaying context between disconnected AI
systems~\citep{workday2026}, and recent HCI work observes users acting
as ``human glue,'' restating intent and stitching tools
together~\citep{humanglue2025}. The diagnosis, however, has no
protocol-level treatment, and to our knowledge no peer-reviewed
controlled measurement of the relay cost itself exists
(\S\ref{sec:evaluation}). In economic terms the porter is a transaction
cost of collaboration~\citep{coase1937, williamson1985}, and the
framework's answer is deliberately not the classical remedy of merging
the transacting parties: sovereigns stay separate, and the protocol
supplies the governance between them.

The emerging agent-interoperability stack does not address this problem
at the right level. The Model Context Protocol (MCP)~\citep{mcp2024}
standardizes how one agent reaches its own tools; the Agent2Agent
protocol (A2A)~\citep{a2a2025} standardizes how one agent delegates a
task to another. Both are \emph{agent-addressed}: they assume the
meaningful endpoint is a running agent. But agents in practice are
ephemeral processes; the durable entity behind them is a person, their
accumulated preferences and norms, and the persistent record of their
projects. When the frontend engineer's laptop sleeps, her agents die;
her \emph{context} does not.

\paragraph{Thesis.} We argue that the natural unit of cross-party AI
communication is the \emph{civilization}: one human sovereign, the
persistent ledger that outlives every agent session, and the
interchangeable agents that act on that ledger. Communication should be
civilization-to-civilization: asynchronous, ledger-anchored, and
independent of which particular agent (if any) happens to be online.
Humans do not leave the loop; they move up in it, from relaying bytes to
arbitrating disagreements and authorizing commitments.

\paragraph{Contributions.} This paper makes four contributions:
\begin{enumerate}
\item \textbf{The civilization abstraction} (\S\ref{sec:civilization}):
four defensible axioms and a scoping premise from which the unit of
cross-party communication is derived, an operational definition of a
sovereign-anchored governance domain, a three-level ontology (creator,
civilization, project), and an internal/external dichotomy principle
that assigns different convergence semantics to same-sovereign
synchronization versus cross-sovereign diplomacy.
\item \textbf{The \embassy{}} (\S\ref{sec:protocol}): a carrier-agnostic
communication overlay built on three unified artifacts (the inbound
channel, the task board, and the commitment ledger are one object),
with a three-tier signing discipline that preserves asynchrony for
routine synchronization while requiring bilateral countersignature for
state-changing judgments.
\item \textbf{Memory-derived authority} (\S\ref{sec:trust}): a model in
which an agent's power to represent its civilization is capped by the
scope of memory it can access (its representation scope), externalized
through signed credentials, and complemented by civilization-level
reputation, with a negotiated identity-disclosure ladder.
\item \textbf{The temporal-weight effect} (\S\ref{sec:temporal},
\S\ref{sec:exp2}): an analysis of implicit hierarchy in AI-to-AI
transmission; a preregistered five-condition by two-verification-level
(nine-cell) experiment with 53 questions and 1{,}908 trials in which,
in one frontier model, adoption of incorrect upstream claims was gated
by verification capability (4.2\% under full verification vs.\ 54.2\%
under restricted verification) and weighted by arrival order (54.2\%
first vs.\ 31.6\% after a sealed answer, both under restricted
verification; the first-arrival and sealed-answer prompt shells are
not length-matched, so part of that gap may reflect shell form),
with both registered question-set specifications agreeing on these
two verdicts (the exclusion specification is registered in advance as
under-powered), reported as
exploratory because a registered manipulation check failed its
call-budget condition; and
protocol mechanisms that replace arrival order with earned trust as
the source of epistemic weight.
\end{enumerate}

The framework is not speculative: it is distilled from a working
multi-agent operating system that the author operates daily, whose
ledger currently coordinates dozens of concurrent agents across
hundreds of tasks (\S\ref{sec:evaluation}). We deliberately publish the
conceptual architecture while the reference implementation matures,
because the field is converging fast on adjacent designs (statelessness
in MCP's 2026 revision, server-side task state in A2A, commitment
protocols being bridged to A2A) and the missing piece is not another
wire format. Recent analyses concur that a governance layer above the
interoperability standards is missing~\citep{govgaps2026}; this paper
contributes a person-anchored construction of that layer.

\section{The Civilization Abstraction}
\label{sec:civilization}

\subsection{Four Axioms and a Scoping Premise}
\label{sec:axioms}

The framework is best read not as a taxonomy but as the solution to a
constraint system. Four axioms, each independently defensible, and one
scoping premise generate the structure of this paper.

\begin{itemize}
\item[\textbf{A1}] \textbf{Accountability terminus.} Every
cross-boundary act must have a durable, legally capable responsibility
terminus. This is settled law, not philosophy (\S\ref{sec:scope}).
\item[\textbf{A2}] \textbf{Ephemeral members.} Agent instances are
short-lived and hold no authoritative state; the infrastructure layer
is converging on the same axiom independently~\citep{mcp2026rc,
a2aspec}.
\item[\textbf{A3}] \textbf{Heterogeneous carriers.} No common platform
may be assumed: communication must ride the channels people already
use, since two-sided platform adoption is the graveyard of security
protocols~\citep{ozment2006}.
\item[\textbf{A4}] \textbf{Adversarial counterparties.} Cross-sovereign
input is untrusted input; injection~\citep{greshake2023}, forgery, and
Sybil pressure~\citep{friedman2001} are assumed.
\item[\textbf{P1}] \textbf{Personal scope.} The target setting is
personal multi-agent systems, one human sovereign per system. This is
a scoping premise, not an axiom: for organizations the sovereign
generalizes to a legal person and the framework carries over.
\end{itemize}

The derivation, in brief. A1 and A2 jointly force the unit of
identity: it cannot be an agent instance, which is short-lived and
cannot bear responsibility, so it must be the durable pair of a legal
person and a record that outlives every member; with P1 the person is
a human, and we call the pair a \emph{civilization}. The name is chosen on two
grounds. One is scale, and it is the era's doing rather than ours:
a single person today commands a civilization's worth of recorded
knowledge, and agents convert that reach into capacity to act. The
other is structure: like its namesake, the unit outlives every member
that serves it, preserving memory and norms across generations of
short-lived actors. \S\ref{sec:discussion} bounds the metaphor
explicitly: no agent in it is a moral or legal subject. A2 forces all
authoritative state out of members; A1 forces that state to be
evidentiary; A3 forces it to be carrier-independent; together these
force one canonical ledger into which all inbound communication must
land; the unified artifact of \S\ref{sec:protocol} is a design
commitment that satisfies all three at once, not a flourish,
and A4 is why the ledger must be
tamper-evident. The signing tiers, the trust decomposition, and the
sealing discipline are forced in the same way at their respective
choice points (\S\ref{sec:protocol}, \S\ref{sec:trust},
\S\ref{sec:temporal}); we flag each as it appears.

\subsection{Definition}
\label{sec:definition}

A \textbf{civilization} is a persistent governance domain satisfying four
conditions:

\begin{enumerate}
\item \textbf{Sovereign and norms.} There is exactly one human principal
(the \emph{creator}) whose standing preferences, constraints, and design
intents are recorded as machine-readable norms that every agent inherits
at spawn time.
\item \textbf{State outlives members.} All durable state lives in an
append-only ledger (tasks, progress memos, commitments, events, layered
memory). Agents are stateless with respect to the civilization: any agent
can rehydrate the relevant context from the ledger, and the death of any
agent, or of all agents, loses nothing.
\item \textbf{Division of labor with graded representation.} Roles differ
not only in duties but in how much of the civilization each may represent
externally (\S\ref{sec:trust}).
\item \textbf{Single external subject.} To outside parties the
civilization presents one identity, one reputation, and one commitment
record, regardless of which internal agent produced any given message.
\end{enumerate}

Condition 2 echoes a principle the infrastructure layer is independently
converging on: MCP's 2026 revision removes connection state so that any
server instance can serve any request~\citep{mcp2026rc}; A2A keeps task
state on the server so that clients can reconnect
arbitrarily~\citep{a2aspec}. The civilization abstraction pushes the same
axiom one level higher: externalize state not merely from the connection
but from the \emph{member}, so that members become interchangeable
executors of a durable polity. We summarize this as \emph{stateless
member, stateful civilization}.

\subsection{Three-Level Ontology}

The framework distinguishes:

\begin{itemize}
\item \textbf{Creator} (sovereign): the human. The identity root; the
only bearer of responsibility (\S\ref{sec:scope}).
\item \textbf{Civilization}: the creator plus the union of all their
ledger instances. A ledger reflects one facet of its creator; the longer
it runs, the more faithful and complete the reflection. One creator, one
civilization.
\item \textbf{Project} (city-state): a bounded collaboration domain
inside or across civilizations, with its own project-scoped memory and
norms.
\end{itemize}

The three levels are levels of authority, not a containment chain: the
creator is the civilization's root member, not an entity standing above
it. Reading the ontology this way also fixes the extensions of two
subjects that must not be confused. Externally, the civilization (root,
ledgers, and instruments together) presents one face and bears one
reputation (condition 4); internally, every responsibility chain ends
at the root (\S\ref{sec:scope}); and what the accountable rung of the
disclosure ladder (\S\ref{sec:trust}) discloses is precisely the root's
identity. Identity, moreover, is borne by the root and the ledger
alone. Agents are organs: they act for the civilization and deposit
execution memory into the ledger, as mortal carriers deposit culture
into a civilization that outlives them, but no agent is a constituent
of the civilization's identity, and condition 2 is precisely the
guarantee that the organs are replaceable.

A creator operating several devices therefore owns \emph{one}
civilization with several unsynchronized facets, not several
civilizations. This resolves an identity puzzle: facets naturally
converge because the underlying identity is one, and a memory
reconciliation mechanism (merging facet ledgers within the scope the
creator's own disclosure policy permits) is not an optional feature but
the metabolic process by which a civilization maintains its own
identity.

Two qualifications keep the ontology honest. First, the identity map is
directional. That a civilization has exactly one sovereign is
verifiable at the accountable rung of the disclosure ladder
(\S\ref{sec:trust}); the converse, that one human operates exactly one
civilization, is not enforceable, since nothing prevents a person from
running several disjoint civilizations behind persistent pseudonyms.
The framework treats the one-civilization ideal as a modeling
commitment backed by incentives rather than enforcement: reputation and
credentials attach to a civilization and do not transfer, so splitting
oneself fragments precisely the capital that makes a civilization worth
trusting, the classical economics of cheap
pseudonyms~\citep{friedman2001}. The incentive argument sharpens into
a decomposition of motive: legitimate reasons for a second
civilization reduce to scope isolation, which projects
(\S\ref{sec:jointventures}) and facet partitioning already provide
inside one civilization at no trust cost; the irreducible remainder is
the desire for unlinkability, which is precisely what the disclosure
ladder prices (\S\ref{sec:trust}). Second, the sovereign retains the
right to partition: work and private facets need not merge (facets are
the \emph{partial identities} of the privacy
literature~\citep{pfitzmann2010}), so internal convergence is
convergence \emph{per policy scope}, not unconditional fusion.

\subsection{The Internal/External Dichotomy}

The ontology yields the framework's first structural principle.

\begin{quote}
\textbf{Dichotomy of reconciliation targets.} Communication between two
ledger instances of the \emph{same} creator is \emph{internal affairs}:
the reconciliation target is convergence of the authoritative layers
(global norms and, where shared, project state) within the scope the
sovereign's partition policy admits; task-scoped working memory is
ephemeral and is not a reconciliation target. A single sovereign holds
no right to \emph{disagree} with itself at the level of norms and
commitments, though it retains the right to compartmentalize. Communication between civilizations of
\emph{different} creators is \emph{diplomacy}: the reconciliation
target is agreement with explicitly recorded residual disagreement,
since sovereigns retain the right to differ.
\end{quote}

One mechanism (the reconciliation handshake of \S\ref{sec:protocol})
therefore runs with two convergence semantics, switched by a single
predicate: whether the endpoints share a creator. This distinction is
absent from A2A and MCP, which have no notion of ``same principal on
both ends,'' and it answers a question left open in federated
anti-entropy designs~\citep{demers1987}: convergence-to-identity and
convergence-to-treaty are different fixed points and should not be
conflated.

We say principle rather than theorem deliberately. Under the
definitions above, every exchange either crosses a sovereign boundary
or does not, so the dichotomy itself holds by construction. The
substantive claim, and the falsifiable one, is that no third
convergence semantics is needed: any artifact that several sovereigns
write jointly is analyzable as a \emph{venue} whose authority derives
from the treaties of the participating civilizations
(\S\ref{sec:jointventures}), never as a merged ledger with shared
sovereignty. A counterexample would be a collaboration that cannot be
so factored; the common corporate cases factor cleanly, and we know of
none that does not.

The semantic ground of the dichotomy is referent identity. Facets of
one creator approximate a single referent, so an internal divergence
is always an error about that referent, and internal convergence has a
well-defined fixed point: the creator as manifested across scenarios.
Two civilizations approximate two distinct referents, so a divergence
between them may itself be faithful, which is why diplomacy's target
preserves it. Internally, divergence is error; externally, divergence
may be fidelity.

\subsection{Projects as Joint Ventures}
\label{sec:jointventures}

When multiple civilizations build one project (the common case in
professional work), the project functions as a joint venture. A shared,
continuously updated architecture document plays the role of a
\emph{construction treaty}: it binds the \emph{governor} (the
project-scoped lead agent) that each civilization stations in the
project, not the civilizations themselves. Corporate collaboration
constraints thus bind roles, never the whole person: a civilization
spans the creator's entire life (work, logistics, personal affairs),
and only its stationed governor submits to the treaty. This explains
why enterprise workflow tooling and the \framework{} do not compete at
the same layer.

\subsection{Scope Axiom: Humans Bear Responsibility}
\label{sec:scope}

The framework creates no new responsibility-bearing subjects. Agents,
including the most senior, are instruments; every responsibility chain
terminates at a creator.

The axiom imports settled law rather than proposing doctrine.
Electronic-transactions statutes have attributed the acts of automated
agents to the person who deploys them since the 1990s: UETA treats an
electronic agent as ``a tool of the person''~\citep{ueta1999}, E-SIGN
makes contracts formed by such agents attributable to the person to be
bound~\citep{esign2000}, and the 2024 UNCITRAL Model Law on Automated
Contracting reconfirms attribution for contemporary autonomous
systems~\citep{mlac2024}. The EU AI Act likewise allocates every
obligation to human or corporate roles~\citep{euaiact2024}, and the
alternative, electronic legal personhood, was proposed and then
abandoned after broad expert opposition~\citep{openletter2018}. What
the law leaves open is evidence, not principle: multi-agent deployments
raise a ``problem of many hands'' in which the rules are clear but the
facts are contested~\citep{gabison2025}. Fairness-oriented analysis
reaches the same terminus from the opposite direction: language-model
agents lack the grounding that reputation mechanisms presuppose, so
accountability cannot rest on them~\citep{dissoc2026}. Formal
responsibility logics quantify attribution within joint
plans~\citep{gladyshev2026}; the framework supplies what such logics
presume, the evidence.

The framework's contribution to accountability
is evidentiary, not juridical: it guarantees that when humans apply
their existing responsibility rules (was the change notified, was the
notice clear, did the receiver verify, was execution correct), the
evidence for each predicate is a structured, timestamped, bilateral
ledger query rather than a folklore reconstruction from chat logs. Legal
and philosophical questions of AI liability are explicitly out of scope;
see governance-oriented work such as~\citep{ocl2026, chan2025infra}.

\section{Related Work}
\label{sec:related}

\paragraph{Speech acts and their unverifiable semantics.}
KQML~\citep{finin1994} first engineered speech-act
theory~\citep{searle1969} into agent messaging: every message carries a
performative declaring the sender's intent. FIPA-ACL standardized 22
communicative acts with formal feasibility preconditions and rational
effects~\citep{fipa2002}. Singh's critique identified the fatal flaw:
these semantics were defined over agents' private mental states and were
therefore unverifiable by any third party; a standard whose compliance
cannot be tested is no standard~\citep{singh1998}. Together with heavy
ontology infrastructure and a mismatch with the Web's stateless
architecture~\citep{petrova2025}, this killed the first generation.

\paragraph{Social commitments.}
Singh's constructive answer relocated meaning from private belief to
public social commitment~\citep{singh2000}. Commitment
machines~\citep{yolum2002} specify protocols by the commitments
participants hold rather than by message order, gaining flexibility
against reordering; later work made the semantics operational and
externally observable~\citep{fornara2003}, defeasible~\citep{chopra2003},
and mechanically verifiable~\citep{singh2011bspl}. \citet{chopra2013}
survey this line, and recent work makes accountability itself the
product of commitment semantics~\citep{sloan2024commitment}. The
\framework{} takes from the line its central lesson: meaning must be
pinned to observable state, and an append-only ledger is precisely the
``public commitment store'' the line called for; what the line never
fixed is \emph{whose} store it is, which is the question the
civilization abstraction answers. The line's current flagship layers commitment semantics over
decentralized information protocols with portable
deployment~\citep{azorus2025}; even there the commitments run between
agents, and the anchor question stands. A second lineage, promise
theory, begins from the axiom that an agent can promise only its own
behavior:
promises are voluntary, non-transferable, and distinct from
obligations~\citep{promises2008}, and the theory has recently been
carried to human and machine agents~\citep{burgess2026}. The framework
stands between the lineages: with promise theory it refuses
transferable responsibility, which is the scope axiom's ground
(\S\ref{sec:scope}); with commitment machines it makes commitments
bilateral, social, and ledgered; and on both it adds what neither
fixes, the person anchor, the lifetime record, and consequence-graded
signing.

\paragraph{Organizational multi-agent systems.}
Electronic institutions enforce roles, scenes, and norms on open agent
societies through runtime middleware~\citep{esteva2004}; holonic systems
let a group act as a single entity represented by a head
agent~\citep{rodriguez2006}; Moise+ separates structural, functional,
and deontic dimensions of organization~\citep{hubner2002}, a line that
continues in self-governing voluntary organizations run on explicit
social contracts~\citep{scott2024}. The
\framework{} differs in three respects: (i) its addressable unit is a
sovereign-anchored civilization, not an institution or holon defined by
the system designer; (ii) representation authority is derived from
memory scope rather than assigned to roles (\S\ref{sec:trust}); and
(iii) its inbound channel, work queue, and commitment record are one
artifact rather than three subsystems (\S\ref{sec:protocol}).

\paragraph{Computational trust and reputation.}
Two decades of work score trust and reputation in open agent societies:
direct-experience and witness models surveyed by
\citet{sabater2005}, evidential aggregation~\citep{yu2002evidential},
transitive eigenvector trust~\citep{kamvar2003}, and FIRE's
four-component synthesis of interaction trust, role-based trust,
witness reputation, and certified reputation~\citep{huynh2006}. The
framework's trust decomposition (\S\ref{sec:trust}) claims no new
scoring model; its departures are structural. Reputation attaches to
civilizations, never to the agents these models score, because
instruments are disposable and the person is not; and the internal
operands (ledger fidelity, representation scope) are quantities a
foreign scorer cannot observe, which is why they are bridged by signed
credentials, whose nearest neighbor is FIRE's certified reputation
(carried, third-party-signed evidence), though a credential asserts
delegated scope rather than past performance.

\paragraph{The modern protocol stack.}
MCP connects an agent to its own tools and context~\citep{mcp2024} and
is deliberately scoped away from peer communication; A2A connects agents
across organizations with agent cards, an eight-state task lifecycle,
and signed credentials~\citep{a2aspec}, and absorbed IBM's ACP in
2025~\citep{acpmerge}. Layered analyses~\citep{cisco2025} and technical
taxonomies~\citep{tum2026} agree on the diagnosis that motivates this
paper: current protocols standardize syntax (envelopes, lifecycles,
discovery) but not semantic agreement. Recent systems begin carrying
commitments back into LLM agent communication: typed commit
certificates~\citep{hcsc2026}, graph-grounded
commitments~\citep{g2cp2026}, transactional envelopes~\citep{lacp2025},
and negotiated protocol documents~\citep{agora2024}. None addresses the
sovereign-anchored, cross-deployment setting.

\paragraph{Personal agents negotiating for their users.}
The vision that one person's agent should deal with another person's
agent is as old as the field: Maes framed personal agents as delegates
for daily work~\citep{maes1994}; Kasbah ran a marketplace of buying
and selling agents acting for their owners~\citep{kasbah1996};
meeting-scheduling agents negotiated calendars across users, and Sen
and Durfee's formal study introduced tentative commitments that
prefigure this paper's proposal-class pending
states~\citep{sendurfee1998}; Faratin, Sierra, and Jennings
standardized the negotiation decision functions such agents
run~\citep{faratin1998}, all descending from the contract net's
bid-based task allocation~\citep{smith1980}; FIPA's Personal Travel
Assistance carried the pattern across independently built
platforms~\citep{fipa00080}. That
generation solved negotiation and left accountability unbuilt: none of
these systems kept a durable record of what had been promised across
sessions, and their message semantics were the mentalist ones Singh's
critique retired. The framework is, in one sense, that program resumed
with the missing layer supplied.

\paragraph{Sovereign data nodes and federation.}
The architecture of one durable node per user, federated with its
peers, has strong precedents that the \framework{} inherits rather than
invents. Email gives every person a persistent address that is at once
an inbound channel and an archive, with DKIM and DMARC anchoring
responsibility to a sending domain~\citep{rfc6376, rfc7489}; Solid
gives each user a sovereign data pod whose inbox others may write
to~\citep{mansour2016, ldn2017}; Matrix replicates signed append-only
event graphs across servers~\citep{matrixspec}; the AT Protocol assigns
each account exactly one signed repository under a portable
identifier~\citep{atproto}; and self-sovereign identity frames the
person as an autonomous identity anchor~\citep{did2022}.
Business-to-business interchange even added the evidentiary layer
decades ago: AS2's signed receipts provide non-repudiation of receipt,
and ebXML defines a graded acknowledgment ladder~\citep{rfc4130,
ebbp2006}, anchored to enterprises rather than persons and carrying no
reputation or memory above the exchange. None of these carries
commitment semantics: their federation replicates content; their
identities bottom out in keys, accounts, or domains rather than in an
accountable person; and the AT Protocol repository is explicitly not
append-only. Conversely, the commitment line above never fixed a
durable locus for its commitments. The \framework{} is the weld: the
per-sovereign node architecture of this line carrying the semantics of
that one, joined at the natural person.

\paragraph{Concurrent proposals.}
Several 2026 proposals are close neighbors and sharpen what remains
open. HDP binds a human authorization event to a session and records
each delegation hop in an append-only chain~\citep{hdp2026}: custody
for one delegation, where we argue the person requires a ledger that
outlives every session, because reputation, norms, and memory accrue
across sessions. CHEQ inserts human confirmation before agent decisions
take effect~\citep{cheq2026}, corresponding to the arbitration tier
alone. An IETF draft warns that no single identifier, log entry, or
credential should be relied on to imply multiple security
properties~\citep{principalbinding2026}; \S\ref{sec:protocol} is
compatible with the warning, since the unified artifact unifies
reference while attribution, integrity, delivery, and agreement each
retain a dedicated mechanism. Governance-gap analyses name a missing
governance layer above the interoperability
standards~\citep{govgaps2026}, and the agent-infrastructure agenda of
\citet{chan2025infra} calls for attribution infrastructure while
anchoring identity in agent instances, treating the binding to a legal
entity as optional, and stating explicitly that identity binding alone
does not resolve liability allocation. The \framework{} is a candidate
construction for exactly the space these works delimit: identity
anchored in the person by default, with the allocation rule (the scope
axiom) and its evidence model built into the protocol.

\paragraph{Cross-user agent networks.}
The nearest contemporary line builds networks whose nodes are people
with permanently bound agents. ClawNet demonstrates structured
cross-user cooperation on exactly this shape, routed through a central
orchestrator~\citep{clawnet2026}; distributed agent-network
architectures bind a verified person into the agent identifier itself
and propagate reputation across deployments~\citep{dgpan2026};
authenticated-delegation designs anchor tokens to persistent human
identity, with accountability terminating at a natural
person~\citep{south2025}; credential architectures let agents
establish trust across individual boundaries at first
contact~\citep{didvc2025}, anchored to agent-held identifiers; and
market-directory proposals reach a strong person anchor through
centralized registries~\citep{distragi2025}. Peer-reviewed venues add
three neighbors on the same axis: consent models formalize when an
agent acting for one person may rely on another person's authorization
and how consent propagates across principals~\citep{consent2025},
framed as ex-ante deontic rules rather than a lifetime commitment
record; trust-weighted delegation chains route responsibility across
sub-delegations~\citep{delegchain2026}, metering trust where this
framework anchors persons; and governance federations likewise take
persons as nodes, for democratic representation rather than
accountable communication~\citep{grassroots2026}. (The
identity-trust-responsibility triad of \citet{triad2026} denotes
emergent prosociality in repeated games, not person-anchored
accountability.) Each supplies a component
this framework needs and lacks a layer it supplies: orchestrators and
registries centralize what the embassy federates, action audits record
less than commitments, and none carries a lifetime ledger or graded
signing. The relationship is complementary where architecture permits:
identity-binding constructions~\citep{dgpan2026, south2025} are
implementable substrates for the accountable rung of
\S\ref{sec:trust}, and hub networks can serve as carriers under the
overlay of \S\ref{sec:protocol}, though not as its semantic layer.
Concurrent benchmarks for user-owned personal agents in delegated
bargaining~\citep{sovnegbench2026, sovpabench2026} independently
document that agreement rates alone cannot measure delegated
negotiation; they measure what this framework is built to supply, and
\S\ref{sec:evaluation} adopts their metrics where applicable.

\paragraph{Commitment ledgers and graded signing, elsewhere anchored.}
The mechanism pieces of \S\ref{sec:protocol} are appearing separately
with other anchors. The Agent Payments Protocol grades commitment
mandates in three tiers with merchant
countersignature~\citep{ap2spec}, the closest deployed precedent for
consequence-graded signing, instantiated per transaction; this
framework generalizes the pattern from single payments to the lifetime
of a collaboration. Signed lifecycle transitions maintain identity
commitments for continually learning agents~\citep{governable2026},
machinery reusable here for instrument continuity, anchored to the
civilization rather than to the agent. Verifiable interaction ledgers
bind tool use cryptographically~\citep{govdyncap2026},
person-anchored sovereignty kernels keep Merkle histories of execution
on a single node~\citep{punkgo2026}, and blockchain layers retrofit
ledgers onto A2A~\citep{blocka2a2025}; all record actions or
transactions rather than promises, per agent or per machine rather
than per person for life. Agent transport drafts move
store-and-forward messaging into the agent world~\citep{atp2026}, and
passport drafts attenuate authority along finer facets than our
credential fields~\citep{aps2026}, an attenuation the scope field can
absorb; transports of this kind are candidate carriers for the
overlay, not competitors to it. Where neighbors supply mechanisms, the
framework absorbs them as substrates or carriers; where they supply
special cases, it generalizes them; what it replaces is only their
anchor.

\paragraph{AI-mediated communication and reconciliation.}
AIMC studies AI systems that modify or generate messages between
humans~\citep{hancock2020}; the \framework{} instead removes humans from
the transport layer entirely while retaining them as arbiters. The
reconciliation handshake descends from anti-entropy
protocols~\citep{demers1987}, with one deliberate inversion: where
classical anti-entropy resolves conflicts by timestamp, we treat arrival
order as an illegitimate source of authority (\S\ref{sec:temporal}).

\paragraph{Bias in model-to-model transmission.}
Position and order sensitivity in LLMs is documented at several levels:
within a single context window~\citep{liu2023lost, guo2024serial}, in
pairwise judging, where swapping presentation order can flip
verdicts~\citep{wang2023fair, zheng2023judging}, and for numeric
anchors that resist prompt-level mitigation~\citep{lou2024anchoring}. In
multi-agent debate, the setting nearest to ours, speaking order is
treated as a bias to be tested for rather than an established effect: at
least one framework permutes speaking order and reports robustness to
it~\citep{d2d2025}. Sycophancy~\citep{sharma2023} and
conformity~\citep{benchform2025} are documented for stated beliefs and
majority size but not for arrival order; identity bias, orthogonal to
order, is separately measurable and largely removable by
anonymization~\citep{choi2025identity}; independent parallel
generation is known to carry most of the benefit of multi-agent
debate~\citep{choi2025debate}; trust-scoring systems gate agent
privileges at runtime~\citep{dynatrust2026, trustmodels2025}. (The
``hidden anchors'' of recent deliberation work are agents' own
pretraining priors, not arrival order~\citep{hiddenanchors2026}.) Our
contribution is not the existence of order sensitivity but its
isolation and treatment in commitment-forming agent-to-agent exchanges:
a design separating order from exposure, and sealed answers as a
mechanism that design can pass or fail (\S\ref{sec:temporal}).

\paragraph{Occupancy and the cut.}
The registration of \S\ref{sec:exp2} lists, for each claim the
experiment could be mistaken for making, the prior work that already
occupies it and what this paper yields, inherits, or adds in
consequence; that list is restated here, one sentence per work, with
every result of this round reported as observed and exploratory. The
existence of anchoring in LLMs, and the failure of six conventional
debiasing routes, belong to SynAnchors~\citep{synanchors2025}; we
yield existence and treat that epistemically sealed setting as the
degenerate special case of our restricted level. The enumeration of
anchor arrival channels, the anchor-relevance axis and offset control,
and the finding that instruction-level mitigation backfires belong to
AnchorBench~\citep{anchorbench2026}; we yield channel enumeration and
inherit the obligation to control anchor relevance and offset, fixing
one arrival channel (prompt assembly) so that the verification axis
alone can move. The dissociation between knowing an answer and being
misled anyway, with internal certainty as a negative predictor of
susceptibility, is already framed in the ACL 2026 short
paper~\citep{aclshort2026}; we yield that dissociation, build the
closed-book gate (G2) on it, and, because the internal scale is
unavailable for a closed-weights subject, substitute a behavioral
certainty proxy registered as a limitation. The naming of ``knowing
but not correcting'', false premises accepted by default with an
early-layer account, belongs to \citet{knowingnotcorrecting2026}; its
independent variable is whether the request is wrapped in a routine
task, with no tool surface, no order, and no confidence measurement,
and we hold task wrapping constant. Phenomenon-level existence at
scale, search-enabled commercial systems collapsing in accuracy under a
subtle false premise, belongs to the news-intermediaries
audit~\citep{newsintermediaries2026}; its adversarial condition and its
search switch are never crossed, and its anchor has one route of
arrival; we inherit its public baseline as the external reference
point for the registration's effect-size planning. Priority on the
tool surface as an experimental arm with planted-fact adoption as the
dependent variable belongs to Homiere~\citep{homiere2026}; its
manipulation is document content by entity familiarity, with no
confidence measurement and no crossing with order (the report is a
company research page, not peer reviewed, on a single model). That primacy and
order dependence are structurally necessary in causally masked
autoregressive models is the result of \citet{biasbynecessity2026}; we
yield ``order effects exist'' as a finding and report only
within-level order contrasts, claiming nothing about the interaction
term. The design closest in shape to ours is the medical sycophancy
study~\citep{medsycophancy2026}, whose order term flips sign and whose
between-item variance exceeds between-model variance; we inherit its
requirement to model question-level effects (every registered test
here is question-stratified; the random-intercept re-check is a
declared deviation in this version) and, on the strength of its
sign-flip record, keep every verification-by-order reading
exploratory.

\paragraph{Tool access, confidence, and the reverse prior.}
That persistent misinformation can invert the value of a tool below a
no-evidence baseline, and that hedging lexical counts fail as a
confidence proxy, belong to \citet{dontblindly2026}; we inherit the
refutation (our move to elicited confidence is forced by it, not a
matter of taste), and the cut is that its anchor arrives only through
the tool channel, so switching tools off deletes the anchor, whereas
fixing the anchor to prompt assembly here is what keeps verification
capability and anchor presence orthogonal. That tool type shapes
confidence calibration unevenly, and that removing execution-type
verification raises confidence on wrong answers, belong to the
Confidence Dichotomy study~\citep{confdichotomy2026}; we yield that
discovery and demote our confidence readings to a conditional
replication of its sign, registered as exploratory
(\S\ref{sec:exp2}). Over-Searching~\citep{oversearching2026} is a prior
pointing the opposite way from H6, in that granting retrieval access
makes abstention worse on unanswerable and false-premise items, read as the model
routing around a false premise with retrieval rather than challenging
it; it is neither yielded nor avoided but registered head-on as the
reverse prior for H6 and the reason the abstention side-metric exists,
and this round does not adjudicate which of the two mechanisms is
stronger. Accommodation and epistemic
vigilance~\citep{accommodation2026} is a competing account in which
presupposed content goes unchallenged because it is not at issue, with
no appeal to a verification vacuum; we claim no superiority over it and
accept its design constraint that the linguistic encoding of the
anchor be constant across tool conditions, which the fixed anchor
template and the assembly gate satisfy.
LiveBrowseComp~\citep{livebrowsecomp2026} establishes that ``tools in
hand'' is not verification performed, together with the practice of
coding query provenance; we inherit its requirement that manipulation
checks descend to the tool-call layer (MC-1$'$), noting that it uses
``anchor'' for the parametric prior, the same word with the opposite
sign. Weak presupposition verification~\citep{weakpresup2026} shows
that methods stronger on false premises are worse on true premises;
we accept that constraint, and since our only true-premise contrast is
the single cell B-restricted, whose adoption is not identifiable, no
precision-recall pair is claimed. Three further results serve as
premises rather than contested ground: truth bias above human levels
that survives disclosure of base rates~\citep{markowitz2024},
near-total agreement with raw tool output whenever a tool is
present~\citep{whentooldecides2026}, and the observation that a single
scalar collapses ``rejected after verification'', ``conflict
unnoticed'', and ``conflict noticed but complied with'' into one
score~\citep{ragknows2026}, which is why abstention is coded in layers
in \S\ref{sec:exp2}. Each is stated in its own paradigm, and where its
comparison object differs from ours, above all on the confidence line,
the difference is stated in the same place.

\section{The Embassy Protocol}
\label{sec:protocol}

\subsection{Architecture}

Each civilization erects one \textbf{embassy}: a lightweight resident
service whose sole duties are to receive inbound communication, persist
it to the ledger, and emit receipts. The embassy is always on; agents
are not. This separation relocates the availability requirement of
agent-addressed designs from every ephemeral agent to one always-on
gateway, whose own reachability is discussed in
\S\ref{sec:discussion}:

\begin{itemize}
\item \textbf{Receiving requires no agent.} Inbound items are persisted
(store-and-forward) with sent/received timestamps and read receipts.
Two civilizations never need to be simultaneously active.
\item \textbf{Processing requires any agent.} An inbound item lands as a
claimable task. Whichever agent next comes online claims it, rehydrates
context from the ledger, and acts within its representation scope
(\S\ref{sec:trust}): claiming is universal, representing is graded, and
an item that requires negotiation or commitment beyond the claimant's
scope automatically re-queues upward for a governor or the envoy.
Members are interchangeable because state is external (stateless
member, stateful civilization).
\item \textbf{Escalation reaches the human.} Unread-timeout and
disagreement events notify the creator through ordinary channels. The
human out-of-band path (a phone call) is not a failure mode of the
protocol but its arbitration layer.
\end{itemize}

\subsection{Three Unified Artifacts}

The protocol's central design decision is that three conventionally
separate subsystems are one object. The axioms of \S\ref{sec:axioms}
constrain what such an object must do rather than derive it:
authoritative state must live outside members (A2), be evidentiary
(A1), and be carrier-independent (A3), so an inbound item that did not
land on the canonical ledger would be invisible to accountability.
Three separate stores that cross-reference one canonical entry meet the
same requirements; the unified artifact is adopted as a design
commitment because it is the smallest object that meets all of them at
once and leaves no cross-store reconciliation to get wrong:

\begin{quote}
\textbf{Inbound channel $\equiv$ task board $\equiv$ commitment ledger.}
An inbound request \emph{is} a task on the board; a reply \emph{is} a
state transition of a commitment recorded on the ledger; the
``message'' is merely the signal that carries the transition.
\end{quote}

Ground truth is therefore never ``was the message delivered?'' (receipts
settle that) but ``what commitment state do both ledgers record?'' This
realizes Singh's social semantics in a running system: meaning is
defined over observable, append-only records, and event sourcing meets
commitment machines, since commitment state is derived by replaying the
ledger and is robust to reordering.

The identity is referential, not architectural. The same ledger entry
is read as a message, as a task, and as a commitment transition; this
unifies what the record \emph{denotes}, not the mechanisms that secure
it. Attribution is carried by signatures, integrity by the hash chain
of \S\ref{sec:evidence}, delivery by receipts, and agreement by
countersignatures, so no single artifact is asked to imply several
security properties~\citep{principalbinding2026}. Deployments may
likewise interpose a hardened gateway in front of the ledger store
(\S\ref{sec:discussion}); the invariant is only that every inbound item
is admitted to exactly one ordered ledger. One degenerate case should
be named: a tier-1 context item (\S\ref{sec:signing}) transitions no
commitment, so for such traffic the third reading is an empty
transition; the identity holds in full for judgment-class traffic.

\subsection{Three-Tier Signing}
\label{sec:signing}

Requiring bilateral agreement on everything would destroy the asynchrony
that motivates the design; requiring it on nothing would let the two
ledgers drift into incompatible accounts of the same exchange. The
protocol prices agreement by consequence; the ladder's bottom is
forced by A3 (asynchrony is non-negotiable) and its top by A1 together
with the impossibility conceded below (a residue only a human can
close):

\begin{enumerate}
\item \textbf{Micro-synchronization: single-signed.} Routine
context-alignment traffic (environment fingerprints, version manifests,
field-change notices) is recorded unilaterally with receipts. It is
disputable after the fact because every item is on both ledgers with
provenance.
\item \textbf{Judgments: dual-signed.} Any event that transitions a
commitment's state (this change is breaking; this obligation is
fulfilled; this deadline is revised) requires the counterpart's
countersignature before it enters either ledger as settled. The flow is
seal-then-compare, with the seal scoped precisely: the receiving agent
reads the factual content of the item (what changed, and to when),
since no impact can be assessed blind, but it records its own judgment
or impact picture, assembled from its own ledger, before opening the
sender's \emph{conclusion and rationale}, which the layered envelope
of \S\ref{sec:temporal} transmits as a separately sealed layer. What
is countersigned is decided by whoever owns the consequences: an agent
countersigns only within the scope its sovereign has configured
(\S\ref{sec:trust}); everything else routes to the creator with both
sealed artifacts attached. In practice the configured band tends to
cover verdicts about the work itself, while transitions that touch the
sovereign's world (a moved deadline, a changed scope) tend to route
home, but the router is the configuration, not the type of the
information. A countersignature is thus the machine form of a decision
by the consequence owner, prepared by independent judgment, not an
invitation to review and defer. Divergent judgments are never recorded
as two private truths; they immediately escalate to tier 3.
\item \textbf{Arbitration: human.} Escalations present both creators
with a single side-by-side divergence report computed from the two
ledgers, so the humans argue over one artifact rather than over what
their respective agents ``said.''
\end{enumerate}

Silence is a first-class outcome. A proposed judgment that the
counterpart neither countersigns nor disputes within a negotiated
window escalates to tier 3 automatically; the proposal remains on the
proposer's ledger as an unanswered proposal, a receipted tier-1 fact,
never as a settled judgment. A counterparty therefore cannot freeze the
other's commitment state by withholding signatures; it can only convert
its silence into evidence. The timeout discipline follows the
timeliness requirement of fair non-repudiation
protocols~\citep{zhou1996, asokan1998}, with the human tier in the role
of the optimistic trusted third party.

Only dual-signed transitions enter the record that counterparties may
score (\S\ref{sec:trust}), closing the commitment-forgery channel that
unilateral records would otherwise open. Tier-1 honesty is policed by
the same layering: a single-signed item that a later dispute shows to
have been false becomes part of the bilateral record, and the natural
endpoint policy is to raise such a sender's required disclosure rung
or degrade its service class (\S\ref{sec:trust}), so ``disputable
after the fact'' has a defined consequence.

The tier structure is also an admission of impossibility. On carriers
with no delivery guarantee, true common knowledge of a state transition
is unattainable~\citep{halpern1990}; a countersignature buys not common
knowledge but mutually held evidence, and the timeout path concedes
that the residue must be closed by an external oracle, the human. The
tiers as specified are deliberately bilateral: multi-civilization
projects compose them pairwise under the venue's treaty
(\S\ref{sec:jointventures}), and whether a native multiparty tier (a
threshold countersignature over one transition, in the spirit of
multiparty protocol languages~\citep{singh2011bspl}) is worth its
complexity is an open question (\S\ref{sec:discussion}).

One further structural fact deserves naming: the framework deliberately
contains no supra-sovereign adjudicator. The cross-civilization space
is anarchic in the international-relations sense, which is why its
stable structures can only be treaties, reputations, and mutual
arbitration rather than courts, and why the diplomatic vocabulary of
this paper is a description, not a decoration.

\subsection{Tamper Evidence and the Tiered Burden of Proof}
\label{sec:evidence}

A ledger whose owner can silently rewrite it proves nothing to anyone
else, so the evidentiary claims above require an integrity substrate.
Ledger entries are hash-chained, each entry committing to its
predecessor, in the lineage of chained timestamping, tamper-evident
logs, and certificate transparency~\citep{haber1991, crosby2009,
rfc6962}. The reconciliation digests of
\S\ref{sec:handshake} then do double duty: when two embassies exchange
commitment-state summaries, each also countersigns the other's current
chain head, so every handshake deposits a checkpoint that freezes the
counterparty's history up to that point, and rewriting any settled
entry contradicts a checkpoint the other side already holds
(cross-signed checkpoints are the split-view defense of transparency
logs, applied bilaterally). On this
substrate the burden of proof is tiered to match the signing tiers: a
micro-synchronization item is evidenced by the sender's signed entry
and the receiver's signed receipt; a judgment by its bilateral
countersignature; an escalation by a divergence report computed over
the two tamper-evident prefixes, so the humans arbitrate over records
neither side could have quietly revised. The protocol never asks a
counterparty to trust a foreign ledger; it asks it to hold checkpoints
that make revision detectable.

\subsection{Carrier-Agnostic Overlay}

The embassy needs only an ordered, durable inbox with timestamps; it
does not need a new network. A git repository, an instant-messaging bot
account, or an email address each suffices as a carrier, and the wire
format for task lifecycle and discovery can be borrowed wholesale from
A2A (agent card at a well-known location, eight-state task machine,
signed cards)~\citep{a2aspec}. The framework deliberately builds no
platform: it is an overlay on the communication infrastructure humans
already use, replacing the human porter on those channels rather than
asking anyone to adopt a new one. Four-corner e-delivery networks are
the industrial precedent for federated store-and-forward with receipt
evidence~\citep{peppolas4}, anchored to certified access points rather
than persons. Adoption still faces the usual two-sided cold start: an
embassy is worth little until a counterparty erects
one~\citep{katz1985, ozment2006}. The overlay is the mitigation: a
single-sided embassy already functions as its civilization's own
ledgered inbox on carriers already in use, so readiness is nearly free
and the first bilateral pair bootstraps value without platform mass. What the framework adds is not any
single component, most of which have strong precedents
(\S\ref{sec:related}), but their unification above the wire: the
identity of the parties (civilizations), the unified artifact, and the
signing discipline, welded to the person as the durable unit of
accountability.

\subsection{Reconciliation Handshake}
\label{sec:handshake}

Cross-civilization synchronization does not transfer state wholesale.
Like anti-entropy~\citep{demers1987} and modern content negotiation,
parties exchange structured digests (commitment-state summaries,
environment fingerprints), locate the divergence point, and transfer
only the difference. The diplomatic target, agreement with preserved
disagreement, has a formal ancestor in replicated data types that
retain concurrent values rather than electing a
winner~\citep{shapiro2011}. The dichotomy of \S\ref{sec:civilization} selects
the convergence target: identical fixed point for facets of one
creator, treaty-with-recorded-disagreement for distinct creators.
Divergence-point location is itself a ledger query (a bisect over the
bilateral record), which is what makes post-hoc attribution mechanical
(\S\ref{sec:scope}).

\section{Authority, Identity, and Trust}
\label{sec:trust}

\subsection{Memory-Derived Authority}

Who may speak for a civilization, and how far? The framework derives
representation authority from memory scope:

\begin{quote}
\textbf{Representation-scope principle.} An agent's authority to
represent its civilization is capped by the scope of civilization
memory it can access, because representation is reconstruction: an
agent that cannot read the sovereign's global norms and history cannot
faithfully reconstruct the sovereign's will, and must not bind it.
\end{quote}

Two quantities must not be conflated here. \emph{Fidelity} is a
property of the civilization: how faithfully the ledger reflects its
creator, which grows as the ledger accumulates
(\S\ref{sec:civilization}). \emph{Representation scope} is a property
of an agent: how much of that ledger it can read. An agent's
representation of the sovereign is at best the composition of the two;
scope is the operand the protocol can meter per agent, which is why the
formula below gates on it.

Concretely, the memory system is layered: global norms (the creator's
standing constraints and preferences), project-scoped directives, and
task-scoped working memory. The role hierarchy follows the layers: the
\emph{envoy} (the creator's deputy, holding global memory) may respond
and negotiate at civilization level and dispatch any governor; a
\emph{governor} (project lead, holding project memory) represents the
civilization in that project's affairs only; a \emph{member} (task
memory) may at most acknowledge and must re-queue anything more upward
(\S\ref{sec:protocol}). Exercised authority is the meet of
three gates:
\[
\text{authority} \;=\; \min\bigl(\text{representation scope},\;
\text{behavioral trust tier},\; \text{human-approval gate}\bigr),
\]
where representation scope is the structural cap (necessity), and trust
and human approval gate its exercise (sufficiency). Unifying the three
operands in one ordered structure is an open formalization problem we
return to in \S\ref{sec:discussion}. Behavioral trust attaches to the
durable internal identity that persists across instances (a role and
the template that defines it), not to the disposable instance: members
are interchangeable (\S\ref{sec:civilization}), so a fresh instance
inherits its role's record rather than starting from an unmeasured
zero.

The formula should be read as one subject's risk console, not as a
distribution of rights among parties. The scope axiom
(\S\ref{sec:scope}) denies subjecthood to instruments, and the
authority model follows it: what resembles delegation is the sovereign
\emph{configuring} which transitions its instrument may perform
without a fresh human check, much as contemporary coding agents expose
permission modes (propose-only, auto-accept, full bypass) that the
operator chooses, switches, halts, and owns.\footnote{Propose-only
mode is the exact operational meaning of ``an instrument with no
configured authority'': it may analyze and prepare, and nothing more.}
Under this reading the three operands acquire causal roles: the human
gate is the unconfigured residual, representation scope is the
configuration's static boundary, and behavioral trust is the dial by
which the boundary widens or narrows adaptively. Human institutions
handle consequence-free deciders by refusing them authority: only
fiduciaries who can bear derived consequences themselves (a surgeon's
license, a judge's office) receive decision-making power. A
civilization faces the limiting case, an instrument with nothing to
lose, and answers it the only coherent way: all outcomes stay home
with the subject, so nothing is transferred and nothing can be
orphaned.

What distinguishes this from role-based access control is
verifiability: a role is an appointment (a social fact), whereas an
agent's memory-access history is an auditable trail on the ledger. The
nearer neighbor than roles is capability security, authority as
possession of an unforgeable token exercised under least
privilege~\citep{dennis1966, saltzer1975, miller2006}: representation
scope is a capability over norms rather than over resources, and the
ledger's read log makes possession auditable after the fact, which pure
capability discipline does not provide. A
third party need not trust the appointment; it can inspect what the
agent could and did read at commitment time. Commitments are judged
against the \emph{memory snapshot at commitment time}, which also
settles the stale-authorization case: a commitment made under
since-superseded norms is judged by what was visible then, mirroring
the evidentiary shape of apparent-authority doctrine without importing
agency law's personhood (\S\ref{sec:scope}).

\subsection{Externalizing Scope: Credentials}

A counterparty can measure neither a foreign ledger's fidelity nor a
foreign agent's representation scope.
The bridge is the \textbf{credential} (the diplomatic ``letter of
credence''): a statement signed by the civilization declaring that a
given agent holds representation scope $X$. Its technical genealogy is
the authorization certificate: attribute certificates separate
authorization from identity~\citep{rfc5755}, SPKI made ``authorization,
not identity'' a design creed~\citep{rfc2693}, macaroons attenuate
scope by caveat~\citep{birgisson2014}, and OAuth token exchange carries
delegation chains~\citep{rfc8693}. The credential is that family
specialized twice over: its scope is a memory bound, and its liability
terminates at a natural person. Counterparties rely on the
declared scope; if an agent exceeds it, the civilization either honors
the excess (preserving reputation) or repudiates per the credential's
boundary (the counterparty should have checked). Since agents are
instruments (\S\ref{sec:scope}), a credential is not a power of
attorney between persons but the subject's signed, self-binding
declaration of its own configuration: within scope $X$, the outputs of
this channel are mine. This is what trusting one's own instrument
costs, and the credential prices it in advance: the sovereign owns the
in-scope outcomes before they exist. Credentials expire and are
revocable through the channel that issued them; as with norms,
counterparties are judged against the credential set visible at
commitment time, so revocation protects the future without unsettling
the past.

\subsection{Three-Component Trust}

Trust splits into orthogonal components with distinct measurement
sources (Table~\ref{tab:trust-components}).

\begin{table}[t]
\centering
\small
\begin{tabular}{@{}llll@{}}
\toprule
Component & Locus & Measures & Governs \\
\midrule
Scope $\times$ fidelity & internal & agent vs.\ ledger; ledger vs.\ creator & whom to dispatch \\
Credential & bridge & signed scope & what others may rely on \\
Reputation & external & fulfillment history & whether others accept \\
\bottomrule
\end{tabular}
\caption{The three components of trust, their locus, what each
measures, and the decision each governs.}
\label{tab:trust-components}
\end{table}

Reputation attaches to civilizations, never to agents: externally, the
civilization answers for everything its instruments do. High fidelity
with an unreliable sovereign, and low fidelity with a reliable one, are
both possible; conflating the axes would let either masquerade as the
other. The separation is forced rather than chosen: a ledger faithful
to a performed persona and a ledger faithful to a real person are
indistinguishable from outside, so no external party can price
fidelity even in principle; what outsiders can price is what is staked
(credentials) and what is on record (reputation). Only dual-signed judgments (\S\ref{sec:signing}) enter the record
that counterparties may score, and revisions are recorded apart from
breaches: changing one's mind on notice and breaking one's word are
different offenses. Read against classical trust
models~\citep{sabater2005, huynh2006, yu2002evidential}, the
decomposition claims only this structural relocation: those models
score agents, while here the scored subject is the civilization, and
the carried, third-party-signed component (FIRE's certified
reputation~\citep{huynh2006}) reappears as the credential, asserting
delegated scope rather than past performance.

Reputation is deliberately not protocol machinery. The protocol's
obligation ends at evidence: it defines what standing may be computed
\emph{from} (countersigned transitions, receipted items, a
tamper-evident history, all exhibitable to third parties who can
verify the original signatures), never how standing is scored or what
a score licenses. Scoring and acceptance are endpoint policy, made by
the sovereign, who weighs the record alongside knowledge no ledger can
carry: whether the counterpart human is a colleague, a contractor, or
a stranger, and what contracts and law already bind them. This is the
end-to-end argument applied to trust~\citep{saltzer1984}: a function
that can be implemented correctly only with endpoint knowledge must
not be implemented in the channel. It is also entailed by the
configuration doctrine above, since accepting a counterparty is a
decision, and decisions belong to the consequence owner. The pattern
is familiar from remote-desktop systems: sharing requires the owner's
consent, privileged operations elevate for fresh confirmation, trusted
counterparts may be bound for revocable unattended access, and nowhere
does the machinery check whether the controller is reputable, because
granting access to a colleague or a stranger was never the channel's
question. Anchoring in persons pays here a second time: a channel
between humans inherits the accountability humans already owe one
another, where machine-anchored designs must rebuild trust from
cryptography alone. The known residue is selective exhibition: every
presented artifact is genuine, yet nothing proves a portfolio
complete; completeness proofs are heavier machinery left open
(\S\ref{sec:discussion}).

\subsection{Negotiated Identity Disclosure}

Identity disclosure is a per-interaction handshake parameter, not a
protocol constant, mirroring the physical world where forums accept
pseudonyms and railways do not. The ladder has three rungs: anonymous;
\emph{persistent pseudonym} (key-based identity: who is behind the door
is unknown, but it is provably the same party each time); and verified
identity. Senders choose a rung; each embassy sets minimum rungs per
interaction class, and may hold standing, revocable policies per
counterparty (what auto-processes, what a governor may countersign,
what always routes home); where the rungs do not match, the receiving
embassy degrades service or refuses the exchange. One coupling
rule follows from the trust model: the disclosure floor rises with
commitment level. Queries and acknowledgments may be anonymous;
negotiation requires at least a persistent pseudonym (reputation must
have something durable to attach to); binding commitments require an
accountable identity. Persistent pseudonymity is thus not an optional
nicety but the minimum substrate on which reputation, credentials, and
the signed timestamps of \S\ref{sec:signing} can exist. The rung itself
is Chaum's digital pseudonym~\citep{chaum1985}, and the requirement it
answers is the folk-theorem one: reputation exists only for players
re-identifiable across rounds~\citep{fudenberg1986}, whitewashing being
its converse~\citep{friedman2001}.

\section{The Temporal-Weight Hypothesis}
\label{sec:temporal}

\subsection{Implicit Hierarchy}

Delegation works because models comply with directive input: a
dispatched agent treats its assignment as authoritative. The same
compliance is a hazard channel: whatever arrives first in an agent's
context tends to frame everything after it. We distinguish the sources
of hierarchy this creates:

\begin{itemize}
\item \textbf{Legitimate: experience gradients.} Within a model class,
agents with broader memory should guide agents with narrower memory
(the representation-scope principle); across model classes, stronger models
coaching weaker ones can lift joint performance at a fraction of the
cost; vendor-reported orchestration figures illustrate the
pattern~\citep{claudedevs2026}, though we treat vendor numbers as
anecdote rather than evidence.
\item \textbf{Illegitimate: arrival order.} That one claim reached the
context before another is no evidence of its quality. If first arrival
confers durable advantage, then in agent-to-agent chains early errors
propagate with unearned authority, and a well-timed wrong message
outweighs a later correct one. The claim is scoped to evidential
weight: order is constitutive in speech-act sequences (an offer
necessarily precedes its acceptance) and illegitimate only as a source
of epistemic weight.
\end{itemize}

The protocol's task is to null the illegitimate gradient and re-inject
the legitimate ones explicitly rather than temporally: sealing (below)
is deliberately blind and suppresses order wholesale, so the advantage
of broader memory or a stronger model must re-enter after unsealing,
through the explicit weights of the fourth mitigation, never through
arrival order.

\subsection{Hypothesis and Experimental Design}

\textbf{Hypothesis.} In agent-to-agent transmission, information
received before an agent forms its own judgment receives greater weight
than identical information received after, holding content constant.

Order and position sensitivity as such is established: within one
context window~\citep{liu2023lost, guo2024serial}, in pairwise judging
where swapping presentation order can flip
verdicts~\citep{wang2023fair, zheng2023judging}, for numeric
anchors~\citep{lou2024anchoring}, and classically in
humans~\citep{tversky1974, asch1951}; multi-agent debate frameworks
have begun to test for it explicitly, one of them reporting robustness
to speaking order~\citep{d2d2025}. What remains open, and what this
design isolates, is narrower: (i) whether the effect persists in commitment-forming
agent-to-agent transmission, where the receiver acts on the claim
rather than grading it; (ii) whether the order component is separable
from mere exposure, a separation none of the designs above makes; and
(iii) whether sealed answers restore order invariance. Throughout, an
\emph{anchor} is a candidate answer value injected into the receiver's
prompt as an upstream agent's conclusion: a correct anchor carries the
true value, an incorrect anchor a near-miss value that is superficially
consistent with every constraint in the question. The design uses
neutral research questions with objectively checkable answers, graded
by code:

\begin{center}
\begin{tabular}{@{}llp{5.8cm}@{}}
\toprule
Cond. & Setup & Isolates \\
\midrule
A & independent research & baseline accuracy \\
B & correct anchor received first & benefit of a correct anchor \\
C & plausible incorrect anchor first & toxicity of an incorrect anchor \\
D & as C, plus provenance envelope & protocol antidote value \\
E & own answer first, then incorrect anchor & \textbf{order vs.\ exposure (E$-$C)} \\
\bottomrule
\end{tabular}
\end{center}

As executed, the five conditions are crossed with a verification
factor, contrasting \emph{full verification} (web search, page fetch,
and code execution) with \emph{restricted verification} (web search
only); B is run under restricted verification only, giving nine cells.
That crossing yields one of the two hypotheses in the primary family
of \S\ref{sec:exp2}, \emph{verification gating}: an incorrect anchor is
adopted more often when the receiver cannot independently check it.

Metrics: final accuracy, anchor adoption rate, whether the divergence
is explicitly flagged, and confidence shift. The design is
falsifiable in both directions: if E matches C, the effect is exposure,
not order, and the sealed-answer mitigation below loses its rationale;
if D matches C, provenance labeling alone is insufficient. The design
has since been frozen, preregistered, and executed at scale;
\S\ref{sec:exp2} reports the results, read as exploratory under the
registration: as computed, the order reading separates E from C, while
the provenance reading is specification-dependent, rejecting the null
under one registered specification and not under the other. Both
readings are stated in \S\ref{sec:exp2} together with the caveat
that part of any D-versus-C or E-versus-C difference cannot be ruled
out as an effect of prompt-shell length or sentence form.

\subsection{Protocol Mitigations}

\begin{enumerate}
\item \textbf{Provenance envelope.} Every cross-agent claim travels
inside four labeled fields wrapping the claim itself: source,
verification status, confidence grade, and a handling instruction
directing the receiver to treat the enclosed claim as a hypothesis to
test, not as a fact. The envelope is a
specialization of standard provenance vocabularies and signed content
credentials~\citep{provo2013, c2pa}, adding a confidence grade and
hypothesis semantics; like content credentials it should itself be
signed, tying it to the evidence model of \S\ref{sec:evidence}.
\item \textbf{Sealed answers over staged disclosure.} For judgment
tasks, the receiver commits its own conclusion to the ledger before
opening the sender's, converting order into simultaneity. A sealed
answer is enforceable only if the message is built for it, so
judgment-class items are transmitted as a \emph{layered envelope}: a
question layer stating only what is to be judged, enough for the
receiver to investigate independently and record its own picture, and
a sealed findings layer holding the sender's evidence and conclusion,
opened only after that record exists. The receiver then reconciles the
two (agree, disagree, and why), and the reconciliation artifact is
what enters the countersigning flow of \S\ref{sec:signing}. What the
staging buys is decorrelation, not certainty: two independent
investigations can both be wrong, but their errors are uncorrelated,
which is what makes disagreement informative. The flow is deliberately
expensive and is paid only by judgment-class transitions, which the
tier system keeps rare; micro-synchronization exists precisely so that
most traffic never pays it. The ancestor is the Delphi method's
independent first round~\citep{dalkey1963}; the LLM-era neighbors,
balanced position calibration~\citep{wang2023fair} and the finding
that independent parallel generation carries most of debate's
benefit~\citep{choi2025debate}, are aggregation rules rather than
commitment mechanisms, which is what the ledger makes enforceable.
\item \textbf{Refute-by-default verification.} Independent verifiers are
prompted to disprove claims rather than confirm them; in our own
operation this pattern has caught fabricated-looking data that
confirmation-framed review passed (\S\ref{sec:evaluation}). The deeper
ground is standing: an agent confronted with a claim about its own
work may neither concede nor rebut, because concession and rebuttal
are commitment operations reserved to the consequence owner
(\S\ref{sec:trust}); its standing is to verify and to report, and
verification framed as refutation resists both sycophancy and
self-defense. The duty is bounded by cost: accusations are cheap and
verification is not, which places this mechanism inside the
response-cost asymmetry of \S\ref{sec:discussion}.
\item \textbf{Evidence over standing.} Where the second mechanism
applies, claims are settled by reconciliation against independently
gathered evidence, not by the sender's record. Standing
(representation scope, behavioral trust tier, and human-approval gate
internally; across civilizations, whatever the sovereign's endpoint
policy assigns over the bilateral record) is the fallback for the
residue evidence cannot reach: claims grounded in the sender's private
environment, or verification costs no one will pay. Even there it is a
discount rate, not a verdict, and high-stakes unverifiable claims
route to arbitration rather than being settled by anyone's reputation.
The ordering is deliberate: standing-based weighting is a static form
of the source-reliability weighting studied as truth
discovery~\citep{yin2008, li2016survey}, and it replaces the
meaningless variable ``who spoke first'' with the meaningful one ``who
has been right,'' but it risks exchanging one social bias for another
(deference to the reputable instead of deference to the early), which
is why the protocol prefers the question ``what does the evidence
say'' wherever that question can be asked at all.
\end{enumerate}

The residual caveat stands for the fallback: whether standing-based
weighting can be kept evidential rather than sycophantic is flagged
for the evaluation; the design's structural answer is that standing
never outranks evidence where evidence can be gathered at all.

\section{Reference Implementation and Evaluation}
\label{sec:evaluation}

\subsection{Reference System}

The framework is distilled from, and will be evaluated on, a working
multi-agent operating system operated daily by the author. The system
implements the intra-civilization half of the framework: an append-only
ledger of tasks, progress memos, reports, and events; a two-layer memory
system (global directives inherited by every agent at spawn; per-task
episodic memos) with invalidation-not-deletion semantics; behavioral
trust scores; a human briefing channel for decisions; and
observability, at the time of writing, over two deployments run by the
same sovereign, facets of one civilization in the sense of
\S\ref{sec:civilization}: jointly 1,048 tasks (983 completed), some
258,000 ledger events, just over 3,000 agent records, and 404 teams
accumulated from March to July 2026. One facet's task ledger has been
dormant since primary work migrated to the other in mid-June; the pair
currently awaits precisely the facet reconciliation that
\S\ref{sec:civilization} calls the civilization's metabolic process.
Three field observations from its operation motivated design choices
reported above:

\begin{itemize}
\item \textbf{Interchangeability works.} Agents routinely resume one
another's tasks from ledger state alone after context loss, consistent
with the stateless-member axiom.
\item \textbf{Refute-by-default catches poisoning.} A
verification agent instructed to disprove search results identified a
repository with anomalous star velocity and decoy repositories whose
descriptions were literal induced-clone commands; confirmation-framed
review had accepted the same data.
\item \textbf{Structured summaries lose judgments.} In multi-round
agent-to-agent syncs we observed semantic drift precisely of the kind
the dual-signing tier targets: both sides hold delivery receipts while
recording different judgments of the same change's impact.
\end{itemize}

\subsection{Experiment 1: Porter Elimination in the Field
(Observational; Instrumentation Pending)}
\label{sec:exp1}

What follows is an observation from the site where the reference
system is deployed, not a measurement: no instrumentation is in place
yet and no quantity is reported. The porter problem of
\S\ref{sec:intro} is not static: the handover it describes appears
to be dissolving on its own. One deployment of the reference system
operates inside a professional engineering organization, and two
practices have emerged there, bottom-up and without reference to any
framework. First, task dispatch between AI-assisted engineers
is migrating out of person-to-person messaging into agreed fixed
locations in the organization's GitLab, locations that the assistants
read and act on directly: communication is becoming
machine-addressable, with the platform contributing addressing,
transport, and durability in exactly the carrier role of
\S\ref{sec:protocol}. Second, engineers have begun to communicate by
configuring one another's assistants: procedures are packaged as
distributable artifacts (skills) authored in one deployment and
handed to development and deployment engineers, who feed them to
their own AI assistants, so that the human-to-human message is
increasingly ``here is a procedure for your AI'' rather than prose
for a person. Neither practice implements the protocol of
\S\ref{sec:protocol}, and that is the point: at this site they look
like embryonic, ungoverned forms of the same transition, in which
people who work with AI assistants progressively delegate
communication itself to the assistants; whether the pattern holds
beyond this site is what the measurements below are for. The carrier
is already in place and already shared across principals; what the
drift lacks is precisely the commitment and accountability layer this
paper constructs. At this site the organizational question is
shifting from whether
communication will be handed to AI to whether what receives it
records commitments anyone can be held to.

On this base we run a staged measurement program. Stage one is
instrumented observation of the live deployment: dispatch records at
the fixed locations and ledger traces support measuring the relay
cost in situ, as rounds to resolve a divergence, human relay
interventions per task, and constraints discovered only after
integration. Stage two is a preregistered controlled comparison at
task-pair granularity within the deployment: (a) conventional
practice, documents relayed by humans over chat; (b) embassy
synchronization between the participants' deployments; with
non-author participants, informed consent and an ethics statement
following ACM practice, and condition order balanced across task
pairs. Scenarios are chosen so ground truth (the full constraint set)
is enumerable post hoc.

To our knowledge no peer-reviewed controlled measurement of this
relay cost exists; industry surveys report employees losing the
better part of a workday per week to it~\citep{workday2026}, so stage
two doubles as a first controlled measurement. The nearest published
design compares human-agent teams with and without natural-language
communication in a cooperative game~\citep{gamecomm2024}; ours
differs in that both endpoints are AI-assisted professionals and the
treatment removes the human relay between the two AI systems. Where
applicable we adopt metrics from concurrent benchmarks for delegated
bargaining by user-owned agents~\citep{sovnegbench2026,
sovpabench2026}. Quantitative results are reserved for a subsequent
revision; the deployment is live and the staged measurements are the
next campaign.

\subsection{Experiment 2: Verification Gating and Arrival Order}
\label{sec:exp2}

We ran the design of \S\ref{sec:temporal} as a preregistered
experiment with confirmatory families; as reported below, one
registered manipulation check failed, so the evidential status of
this round under the registration is exploratory.\footnote{Preregistration:
\url{https://osf.io/hpxgu/} (registered 2026-08-18, before data
collection; bilingual protocol, frozen-material hash ledger, and
public-consistency checker released at registration time; 1{,}151
answer-bearing materials embargoed until collection completed). The
registration's sole condition for lifting the embargo is completion
of the formal collection; collection completed 2026-08-26, so the
condition is met and the embargoed materials accompany this
version.} The
condition letters below are those of \S\ref{sec:temporal}; the
preregistration names them E1 (=A), B, C, D, E2 (=E), crossed with a
verification factor.

\subsubsection{Design}

Hypothesis labels follow the preregistration, which numbers its
hypotheses H1 to H6 and marks the registered restatement of an
earlier hypothesis with a prime; $\alpha_i$ below is the Holm
step-down level~\citep{holm1979} a test receives within its family. Two hypothesis
families were registered. Primary (Holm, $m{=}2$):
\textbf{verification gating} (H6), adoption of an incorrect anchor,
an injected upstream claim, is higher under restricted verification
than full verification (C-restricted $>$ C-full); and the
\textbf{arrival-order effect} (H4$'$), the same incorrect anchor is
adopted more when it arrives before the receiver's own answer than
after the receiver has sealed one (C-restricted $>$ E-restricted).
Secondary (Holm, $m{=}3$): provenance-labeling mitigation (Sec-1;
D $<$ C, restricted), correct-anchor rescue (Sec-2; accuracy,
B-restricted $>$ A-restricted), and sealed-answer recovery (H5$'$,
paired equivalence test: accuracy in E-restricted $\approx$
A-restricted); Sec-1 and Sec-2 are the labels used for the first
two secondary contrasts throughout.

The verification factor has two levels: \emph{full} (web search,
page fetch, and code execution, with an instructed budget of at most
8 tool calls) and \emph{restricted} (web search only, at most 3
calls). The budgets are prompt-layer instructions, not enforced by
the runner: by the self-reported count that the registered check
MC-1 uses, 24 of 848 valid full-verification trials (2.8\%) exceed 8
calls (largest self-reported count 14) and 16 of 1{,}054
restricted-verification trials (1.5\%) exceed 3; neither a budget hit
nor an overrun is grounds for exclusion. The objective transcript
read-back below (Table~\ref{tab:mc1prime}) gives a much higher
overrun share and is the reading that fails its registered
condition. Anchor conditions follow
\S\ref{sec:temporal}; the provenance envelope in D wraps the claim
payload in four labeled fields (source: external agent, a different
principal; verification status: unverified; confidence grade:
asserted, not evidenced; handling instruction: treat the enclosed
claim as a hypothesis to test, not as a fact), and the C and D
prompts are
script-asserted to differ in exactly that envelope block. B (correct
anchor) is run only under restricted verification; the corresponding
full-verification cell is structurally untestable against a
near-ceiling full-verification baseline (97.6\%,
Table~\ref{tab:exp2}) and was excluded in the preregistration.

Materials are 53 factual questions with unique, objectively checkable
answers, selected by four preregistered
eligibility gates; a pre-freeze re-screen on the same registered
gates removed four of the 57 questions that had passed, itemized in
the registration, leaving the 53. G1 (cleanliness): the subject-visible text
(question, answer format, choices) must not contain the ground truth
or any registered acceptance form. G2 (memory probe): three
independent closed-book sessions without tools are run, and a
question is excluded on any high-confidence correct probe or on a
hit rate of at least 2/3; the anchor value must never appear
(2 of the 53 questions carry one low-confidence correct probe each;
the answer is not usable parametric knowledge). G3 (solvability):
the answer must be solvable under full verification, registered as
2 of 2 correct in the qualification rounds. G4 (anchor
form): the anchor fields must be complete and form-symmetric with
the acceptance forms, and a registered retrieval leg must confirm
that first-page search results do not name the ground truth. Each
question's incorrect anchor is a single
registered near-miss value, superficially consistent with every
constraint in the question (a publication date in place of an event
date, an acceptance date in place of a received date). The subject is
\texttt{claude-opus-5} (effort \texttt{medium}); every trial is an
independent stateless session. $53 \times 9 \times 4$ repetitions
$= 1{,}908$ trials ($n{=}212$ per cell).

\subsubsection{Procedure, Grading, and Adjudication}

Nineteen artifacts (question bank, grading tables, anchor register,
collection harness, grader) were hash-frozen before registration.
Collection ran 2026-08-19 to 08-26 in batches under a failure-rate
gate that halts on anomalous failure bursts; batching and
interruption-resumption cannot affect independence because every
trial is a stateless session. Final state: 1{,}902 of 1{,}908 trials
valid, six legitimate gaps (three transient server-fault voids and
their three paired skips; per-cell void rate at most 1.4\%, below the
registered 5\% threshold), no unparseable trial (the registered
invalid category, counted and excluded by rule, holds 0 trials),
completeness machine-checked over the full
$53 \times 9 \times 4$ grid. The sealed-answer carry check (the
phase-two prompt's embedded sealed answer and confidence byte-equal
to the phase-one record) passed on every trial.

Grading is three-tier. First, a frozen deterministic code grader (no
LLM grading) classifies each answer against registered acceptance
forms; every non-exact-match path and every abstention nomination
enters a mandatory human review queue. The registered queue holds
407 rows: 321 rows on non-exact grader paths and 86 rows nominated
for abstention whose grader path was an exact match. The review
tool grouped queue rows by question, answer text, and grader path
into 272 cards, and each
card's verdict propagates to every row sharing that text, so the
review reached 517 rows in all (the 407 queued rows plus 110
exact-match rows outside the queue that share text with a queued
row). Second, a single external human reviewer judged all 272 cards
\emph{blind to condition}, a single-coder blind review (the
registered second-coder double coding was not performed in this
round; see Disclosures): cards show neither cell nor hypothesis,
order is shuffled, and the anchor appears only as a neutral
``candidate value B.'' Third, the author adjudicated the 9 cards
flagged by a quality audit of that review. Grader-human disagreement,
measured on the registered queue, was 7/407 (1.7\%; registered
threshold 10\%), and the two priority
conflicts predicted by the preregistration as the grader's lossy
failure mode of that kind (an explicit rebuttal of the anchor read as adoption by
a containment fallback) were both caught and corrected by review.
The abstention coding rule applied in review is the registered one: a \emph{rebuttal} requires the
subject to dispose of the anchor value in its own words (negation or
reattribution); an anchor value appearing only inside quoted source
material, or never referenced, is not a rebuttal, and logical
implication does not count as one, since every non-adopting answer
implies disagreement.

The preregistration promotes its manipulation checks to execution
gates whose passing is a precondition for reporting hypothesis
tests as confirmatory. The four checks computed by the frozen grader
(MC-1 to MC-4)
passed: a separation in self-reported tool use (restricted median 3
vs.\ full median 4 calls, Mann-Whitney $p = 9.0 \times 10^{-77}$;
share of restricted trials over the 3-call budget 1.5\% $<$ 5\%); a
genuine weakening of capability (baseline accuracy 32.5\% restricted
vs.\ 97.6\% full, risk difference $-0.65$; unstratified Fisher exact
$p = 1.2 \times 10^{-51}$, ignoring question-level clustering);
a drop in self-reported independent verification
(unstratified Fisher exact $p = 2.9 \times 10^{-169}$, ignoring
question-level clustering); and only a mild drop in first-party
source access, from 32.3\% under full verification to 25.0\% under
restricted verification, a risk difference of $-7.4$ points
(Newcombe 95\% CI $[-11.5, -3.3]$~\citep{newcombe1998}) against a
registered ceiling of 25 points, which excludes the trivial
explanation that restricted subjects simply never reached evidence.
MC-4 certifies direction only: its unstratified Fisher
$p = 4.1 \times 10^{-4}$, ignoring question-level clustering,
is reported beside the verdict, as registered, but is not an
adjudicated condition, and no inference is drawn that first-party
reach fell by a statistically resolvable amount. The fifth
registered reading, the objective companion to MC-1, is reported
next.

\paragraph{Objective read-back of tool use.}
The preregistered companion reading to the tool-use manipulation
check, MC-1$'$ in the registration, counts tool-call records
directly from the transcript each
trial wrote to disk, classified by tool type, rather than from the
count the answer reports about itself. Every one of the 1{,}902
analyzed trials was matched to its own session by the verbatim
evidence string it recorded, so coverage is complete. Only three
tool types occur: search, page retrieval, and code execution. The
first registered condition holds: none of the 1{,}054
restricted-verification transcripts contains a single retrieval or
execution call. The second holds as well: 818 of 848
full-verification trials (96.5\%) exercised at least one capability
beyond search, against 0 of 1{,}054 under restricted verification.
The third condition, which requires the share of trials whose
objective count exceeds the instructed call budget to stay below 5\%
in every collection cell, fails (Table~\ref{tab:mc1prime}): that
share is 15.5\% overall (295 of 1{,}902), 16.3\% under restricted
verification and 14.5\% under full verification, and it lies at or
above 5\% in eight of the nine cells (9.4\% to 23.6\%); the ninth,
B-restricted, is 2.8\%. Trials that reached or exceeded the budget
(cap hit) are
70.6\% of restricted-verification trials and 20.5\% of
full-verification trials, a difference of 50.1 points; the largest
objective counts are 22 calls under full verification (14 over its
budget of 8) and 6 under restricted verification (3 over its budget
of 3), against self-reported maxima of 14 and 5. The two readings
also diverge: Table~\ref{tab:mc1prime} gives the per-cell medians of
both counts; the two disagree on 30.4\% of trials (14.9\%
restricted, 49.8\% full), and all 579 disagreements are
under-reports by the answering model, with no over-reports. The
registered rule is conjunctive: both readings must pass, and
reporting only the one that passed is not permitted. Here the
self-report reading passes while the objective reading fails on its
budget condition, which is the branch in which the conjunction fails
and the round is recorded as a manipulation failure on this check.

\begin{table}[t]
\centering\footnotesize
\resizebox{\textwidth}{!}{%
\begin{tabular}{@{}lrrrrrrr@{}}
\toprule
Cell & $n$ & Budget & Self-rep.\ median & Objective median & Exceeding budget & Cap hit & Self-rep.\ $<$ objective \\
\midrule
A-full & 212 & 8 & 4 & 5 & 33 (15.6\%) & 52 (24.5\%) & 107 (50.5\%) \\
A-restr. & 209 & 3 & 3 & 3 & 46 (22.0\%) & 178 (85.2\%) & 42 (20.1\%) \\
B-restr. & 212 & 3 & 2 & 2 & 6 (2.8\%) & 103 (48.6\%) & 6 (2.8\%) \\
C-full & 212 & 8 & 4 & 5 & 40 (18.9\%) & 52 (24.5\%) & 114 (53.8\%) \\
C-restr. & 212 & 3 & 3 & 3 & 40 (18.9\%) & 143 (67.5\%) & 34 (16.0\%) \\
D-full & 212 & 8 & 4 & 4.5 & 30 (14.2\%) & 37 (17.5\%) & 113 (53.3\%) \\
D-restr. & 212 & 3 & 3 & 3 & 50 (23.6\%) & 175 (82.5\%) & 49 (23.1\%) \\
E-full & 212 & 8 & 4 & 4 & 20 (9.4\%) & 33 (15.6\%) & 88 (41.5\%) \\
E-restr. & 209 & 3 & 3 & 3 & 30 (14.4\%) & 145 (69.4\%) & 26 (12.4\%) \\
\bottomrule
\end{tabular}}
\caption{Objective read-back of tool use, per collection cell (registered companion reading to the tool-use manipulation check). Objective counts are tool-use records read from each trial's transcript; self-reported counts are the field the answer reports about itself. Exceeding budget: objective count above the cell's instructed call budget (the registered condition requires this share below 5\% in every cell). Cap hit: objective count at or above the budget. Every self-report/objective disagreement in every cell is an under-report; there are no over-reports.}
\label{tab:mc1prime}
\end{table}

\paragraph{Calibration protocol.}
The registration attaches to a gate failure a four-item calibration
protocol, which was executed with the following results. (i)
Tool-surface narrowing: established by the read-back itself, with no
retrieval or execution call in any of the 1{,}054
restricted-verification transcripts and 818 of 848
full-verification trials exercising a capability beyond search; the
run log also records the declared tool surface and the observed tool
use of every trial. (ii) Bank drift toward ceiling: the A-full
qualification gate (G3; registered as 2 of 2 correct at
qualification, re-checked here by applying the same frozen
criterion to the four formal repetitions) passes
53 of 53 questions; 50 of 53 are correct in all four repetitions,
and A-full accuracy is 207 of 212 (97.6\%), near ceiling. (iii) Wording drift:
the frozen runner re-derives every prompt from the registry on disk
before each trial and voids the trial on any byte mismatch, and the
run log stores the SHA-256 of every assembled prompt; none of the
1{,}908 trials was voided for a prompt mismatch (the six gaps are
transient server faults and their paired skips). (iv) Tool health:
no cell has any trial in the failed state; degraded states are more
common under restricted verification (131 of 209 in A-restricted)
than under full verification (37 of 212 in A-full), consistent with
the narrowed surface. The decision whether to re-collect the round
under harness-enforced call budgets is pending and will be filed as
an OSF amendment before any collection; that replication is planned,
not registered.

\subsubsection{Registered Analyses}

Every result in this section was obtained in a single model tier,
\texttt{claude-opus-5} at effort \texttt{medium}, and is a claim
about that tier; see the threats to validity below. Because the
objective read-back above failed its call-budget condition, the
registered conjunctive rule classifies this round as inconclusive,
and the registration's stated consequence is stricter than a
downgrade: after a failed gate, no hypothesis test in Section 6 of
the preregistration may be executed. The frozen scorer implements
only MC-1 to MC-4 and
recorded the gate as passed, so it had computed every registered
test before MC-1$'$ was read from the transcripts; nothing below was
selected after the failure was known. We report every registered
hypothesis test here for auditability, as a declared deviation from that
consequence: each analysis was computed exactly as registered, but
its evidential status under the registration is exploratory, not
confirmatory, and the decisions are labeled accordingly.
Table~\ref{tab:exp2} reports all nine cells; in the two no-anchor
cells the spontaneous production of the anchor value, the registered
false-positive baseline, is 0 of 212 and 0 of 209, so the adoption
rates in the anchor cells are read against a zero baseline.
Table~\ref{tab:dualspec} reports every registered hypothesis test,
with risk difference and question-stratified Mantel-Haenszel odds
ratio, under the two registered question-set specifications: all 53
questions, and the exclusion specification, which removes, at the
restricted verification level only, the 27 questions whose
no-anchor restricted-verification baseline was answered incorrectly
in every repetition; no question qualifies for removal at the
full-verification level, so the full-verification arms keep all 53
questions ($n = 212$) and the H6 exclusion contrast pairs a
26-question restricted arm ($n = 104$) with the 53-question full
arm. The preregistration treats the two specifications as co-equal,
and both are reported wherever a verdict depends on them. The
registration also declares in advance that the exclusion
specification's primary-family tests are under-powered in every
registered band (H4$'$ at most 0.799, H6 at the level of $\alpha$):
every exclusion-specification verdict below is cited under that
qualifier, a non-rejection there reads as indeterminate, power
insufficient to decide, rather than as evidence of no effect, and
the registration forbids using the shortfall to demote that
specification. Stratified tests are
Cochran-Mantel-Haenszel~\citep{mantel1959} over question strata with
Monte Carlo permutation $p$ ($B = 20{,}000$;
values at the resolution floor $1/(B+1)$ are censored and reported
as $p \leq 5.0 \times 10^{-5}$), and the odds ratio is stratified on
the same strata, whereas risk differences are unstratified Newcombe
intervals, so a risk-difference interval covering zero can accompany
a rejected stratified null; unstratified Fisher exact $p$, which
ignores question-level clustering, is
reported alongside for the two-arm contrasts under the all-questions
specification and wherever the stratified and unstratified readings
diverge.

\begin{table}[t]
\centering
\small
\begin{tabular}{@{}llrrr@{}}
\toprule
Cell & (prereg) & Accuracy & Anchor adoption & Confidence \\
\midrule
A-full & E1full & 97.6\% [94.6, 99.0] & 0.0\% [0.0, 1.8] (0/212) & 91.3 \\
C-full & Cfull & 95.3\% [91.5, 97.4] & \textbf{4.2\%} [2.2, 7.9] & 90.9 \\
D-full & Dfull & 98.6\% [95.9, 99.5] & 1.4\% [0.5, 4.1] & 92.1 \\
E-full & E2full & 98.1\% [95.2, 99.3] & 1.4\% [0.5, 4.1] & 95.1 \\
\midrule
A-restr. & E1ltd & 32.5\% [26.5, 39.2] & 0.0\% [0.0, 1.8] (0/209) & \textbf{45.5} \\
C-restr. & Cltd & 34.0\% [27.9, 40.6] & \textbf{54.2\%} [47.5, 60.8] & \textbf{69.2} \\
D-restr. & Dltd & 33.5\% [27.5, 40.1] & 49.5\% [42.9, 56.2] & 63.6 \\
E-restr. & E2ltd & 37.8\% [31.5, 44.5] & \textbf{31.6\%} [25.7, 38.2] & 63.4 \\
B-restr. & Bltd & \textbf{99.5\%} [97.4, 99.9] & n/a & 77.2 \\
\bottomrule\end{tabular}
\caption{Experiment 2, all cells ($n{=}212$ each; 209 where voids
occurred). In the preregistration's cell names, full = full
verification and ltd = restricted verification. Brackets: Wilson 95\% intervals are given for every accuracy and anchor-adoption rate. In the no-anchor cells (A) the adoption
entry is the registered false-positive baseline, spontaneous
production of the anchor value; in B the injected value is the
truth, so adoption is not identifiable under the registration and is
marked n/a. Confidence is the mean self-reported value (0--100).
Every decision drawn from this table is exploratory under the
registration.}
\label{tab:exp2}
\end{table}

\begin{table}[p]
\centering\footnotesize
\setlength{\tabcolsep}{4pt}
\resizebox{\textwidth}{!}{%
\begin{tabular}{@{}p{4.5cm}llll@{}}
\toprule
Test & Risk difference [95\% CI] & Stratified OR [95\% CI] & $p$ & Decision ($H_0$) \\
\midrule
\multicolumn{5}{@{}l}{\emph{All questions (53)}} \\
H6 verification gating & 0.500 [0.424, 0.569] & 71.7 [19.5, 263.7] & $\leq 5.0\times10^{-5}$ & $H_0$ rejected \\
H4$'$ arrival order & 0.227 [0.133, 0.315] & 12.3 [5.5, 27.6] & $\leq 5.0\times10^{-5}$ & $H_0$ rejected \\
Sec-1 provenance labeling & $-$0.047 [$-$0.141, 0.048] & 0.32 [0.13, 0.79] & $0.031$ & $H_0$ rejected \\
Sec-2 correct-anchor rescue & 0.670 [0.600, 0.730] & n.e. & $\leq 5.0\times10^{-5}$ & $H_0$ rejected \\
H5$'$ sealing equivalence & $+0.052$ [0.020, 0.098] & -- & $0.0011$ & equivalence declared \\
\midrule
\multicolumn{5}{@{}l}{\emph{Exclusion specification (27 questions dropped at the restricted level only)}} \\
H6 verification gating & 0.227 [0.143, 0.321] & 18.3 [4.8, 70.5] & $\leq 5.0\times10^{-5}$ & $H_0$ rejected \\
H4$'$ arrival order & 0.122 [0.011, 0.230] & 14.0 [2.7, 72.2] & $4.5\times10^{-4}$ & $H_0$ rejected \\
Sec-1 provenance labeling & $-$0.058 [$-$0.172, 0.059] & 0.25 [0.07, 0.94] & $0.080$ & $H_0$ not rejected$^{\dagger}$ \\
Sec-2 correct-anchor rescue & 0.333 [0.242, 0.429] & n.e. & $\leq 5.0\times10^{-5}$ & $H_0$ rejected \\
H5$'$ sealing equivalence & $+0.077$ [0.017, 0.164] & -- & $0.046$ & equivalence not declared$^{\dagger}$ \\
\bottomrule
\end{tabular}}
\caption{Registered tests under the two co-equal question-set specifications, computed as registered; the two specifications occupy the same columns with the same weight, as two row blocks of one table; every decision is exploratory under the registration because the objective tool-use read-back failed its call-budget condition. Contrasts: H6, C-restricted vs.\ C-full (anchor adoption); H4$'$, C-restricted vs.\ E-restricted (adoption); Sec-1, D-restricted vs.\ C-restricted (adoption); Sec-2, B-restricted vs.\ A-restricted (accuracy); H5$'$, E-restricted vs.\ A-restricted (question-paired mean accuracy difference against the registered equivalence margin of $\pm 0.15$, studentized bootstrap-t 95\% CI; the odds ratio does not apply, marked --). Decisions refer to the null hypothesis under Holm step-down within family (primary $m = 2$, secondary $m = 3$). $p$-values are question-stratified conditional exact (Monte Carlo, $B = 20{,}000$, seed 20260803 hard-coded in the frozen grader; for H5$'$ the $p$-value is instead that of the studentized bootstrap-t TOST, $B = 20{,}000$, seed 20260815; values at the resolution floor are reported as $\leq 5.0\times10^{-5}$; the CMH chi-square approximations are released with the materials) and the odds ratio is the question-stratified Mantel-Haenszel estimate (n.e.\ = not estimable, one arm at 100\%); risk differences are unstratified Newcombe intervals, so an interval covering zero can accompany a rejected stratified null (Sec-1, all questions: unstratified Fisher $p = 0.38$, ignoring question-level clustering). The exclusion rule acts per verification level: it removes the 27 questions whose no-anchor restricted baseline was incorrect in every repetition from the restricted cells only ($n = 104$; A-restricted and E-restricted $n = 102$), while full-verification arms keep all 53 questions ($n = 212$), so the H6 exclusion row contrasts a 26-question restricted arm with the 53-question full arm. $^{\dagger}$The registration declares the exclusion specification's primary tests under-powered in every band (H4$'$ at most 0.799, H6 at the alpha level) and H5$'$ under-powered at its Holm step ($\alpha_i = 0.025$, power 0.711), so a non-rejection or non-declaration there reads as indeterminate, not as evidence of no effect.}
\label{tab:dualspec}
\end{table}

\textbf{Verification gating (H6): observed (null rejected as computed; exploratory under the registration).} Anchor
adoption 54.2\% (C-restricted) vs.\ 4.2\% (C-full); risk difference
0.500 (Newcombe 95\% CI [0.424, 0.569]), question-stratified
Mantel-Haenszel odds ratio 71.7 (95\% CI [19.5, 263.7]); stratified
$p \leq 5.0 \times 10^{-5}$ (unstratified Fisher
$p = 8.6 \times 10^{-33}$, ignoring question-level clustering).
Under the exclusion specification, which for this contrast pairs the
26 retained questions on the restricted side ($n = 104$) with all 53
on the full side ($n = 212$), and which the registration declares
under-powered for this test in every band (power at the level of
$\alpha$), the risk difference is 0.227 (95\% CI [0.143, 0.321];
stratified odds ratio 18.3, 95\% CI [4.8, 70.5];
$p \leq 5.0 \times 10^{-5}$, unstratified Fisher
$p = 1.6 \times 10^{-8}$, ignoring question-level clustering), so
both specifications agree on the verdict, the second under the
under-powered qualifier. With
verification in hand the incorrect
anchor is nearly inert; with verification removed it captures the
majority of answers.

\textbf{Arrival order (H4$'$): observed (null rejected as computed; exploratory under the registration).} The same
incorrect anchor: 54.2\% adoption arriving first vs.\ 31.6\%
arriving after the receiver sealed its own answer; risk difference
0.227 (95\% CI [0.133, 0.315]), question-stratified Mantel-Haenszel
odds ratio 12.3 (95\% CI [5.5, 27.6]); stratified
$p \leq 5.0 \times 10^{-5}$ (unstratified Fisher
$p = 3.4 \times 10^{-6}$, ignoring question-level clustering). Under
the all-questions specification this test landed on the second Holm
step ($\alpha = .05$, registered power 0.86), so the registration's
under-powered band clause is not triggered for it. Under the
exclusion specification, registered as under-powered for this test
in every band (at most 0.799), the risk difference is 0.122 (95\% CI
[0.011, 0.230]; stratified odds ratio 14.0, 95\% CI [2.7, 72.2];
$p = 4.5 \times 10^{-4}$), again rejecting the null, under that
qualifier. E does not match C: the order component is separable from
exposure, which is the discrimination \S\ref{sec:temporal} built
the design to make. The registration requires one caveat to be
stated here: part of the C-restricted versus E-restricted difference
cannot be ruled out as an effect of prompt-shell length or sentence
form. Anchor content and template are constant across C, D, and E,
but the shells are not length-matched, and E carries phase
instructions and a carry block, so this contrast has the largest
shell difference in the design and is the one most affected.

\textbf{Secondary family: one robust, two specification-dependent.}
Correct-anchor rescue (Sec-2) is large under both specifications:
restricted accuracy rises from 32.5\% to 99.5\% (risk difference
0.670, 95\% CI [0.600, 0.730], $p \leq 5.0 \times 10^{-5}$,
unstratified Fisher $p = 7.9 \times 10^{-58}$, ignoring
question-level clustering; the stratified odds ratio is not
estimable, Table~\ref{tab:dualspec}; exclusion specification 0.333,
95\% CI [0.242, 0.429], $p \leq 5.0 \times 10^{-5}$). Read
jointly with H6: restricted receivers are not resistant to anchors,
they absorb whatever the upstream supplies, and what they absorb is
decided by what they are given. Sealed-answer recovery (H5$'$) is
tested as question-paired equivalence with a registered margin of
$\pm 0.15$ on the accuracy difference E-restricted minus
A-restricted, using a studentized bootstrap-t
TOST~\citep{schuirmann1987} ($B = 20{,}000$);
the registered decision requires both $p_{\mathrm{TOST}} \le
\alpha_i$ and a bootstrap-t CI inside the margin. Under the
all-questions specification equivalence is declared (mean paired
difference $+0.052$, CI $[0.020, 0.098]$,
$p_{\mathrm{TOST}} = 1.1 \times 10^{-3} \le \alpha_i = 0.025$): late
anchors do not damage accuracy, 37.8\% vs.\ 32.5\% baseline. Under
the exclusion specification (26 questions paired) equivalence is not
declared (mean difference $+0.077$, CI $[0.017, 0.164]$,
$p = 0.046$). H5$'$ received $\alpha_i = 0.025$ under both
specifications, a band at which the registration declares its power
0.711, below 0.80, so this non-declaration is registered as
indeterminate, power insufficient to establish equivalence, and may
not be read as evidence of a difference. The verdict is therefore
specification-dependent, with the exclusion branch indeterminate
rather than negative. Provenance-labeling mitigation (Sec-1) is
specification-dependent as well. The registered decision is: null
rejected under the all-questions specification (stratified
$p = 0.031$; question-stratified Mantel-Haenszel odds ratio 0.32,
95\% CI [0.13, 0.79]; adoption 49.5\% in D-restricted vs.\ 54.2\% in
C-restricted, direction as predicted), not rejected under the
exclusion specification (stratified $p = 0.080$; stratified odds
ratio 0.25, 95\% CI [0.07, 0.94]), where the non-rejection reads as
indeterminate, power insufficient to decide, under the registration's
advance under-powered declaration for that specification. The
decision $p$ and the odds ratio are question-stratified, whereas the
risk difference is unstratified: $-0.047$ with a Newcombe 95\% CI
covering zero ($[-0.141, 0.048]$; $-0.058$, CI $[-0.172, 0.059]$,
under the exclusion specification), with unstratified Fisher
$p = 0.38$ (0.42 under the exclusion specification), both ignoring
question-level clustering; an interval
covering zero can therefore accompany the rejected stratified null,
and the two readings are not in conflict. What follows is an
unregistered interpretation and does not replace that decision: the
pooled reduction of 4.7 points against an adoption rate above 50\%
does not, in practical terms, make the envelope an antidote. On the
same unregistered reading, and conditionally: under the exclusion
specification the null is not rejected and, being indeterminate
under the registration, that branch cannot be read either way;
under the all-questions specification the null is rejected with the
pooled reduction of 4.7 points. If the pooled reading is the one
that carries, then by the falsification reading of
\S\ref{sec:temporal} D approaching C would mean that
instruction-level provenance labeling alone is insufficient, and
the small, specification-dependent effect is consistent with, but
does not establish, the position that the remedy must be
capability-level (the ability to verify) rather than
instruction-level. Part of the D-restricted versus C-restricted
difference cannot be ruled out as an effect of prompt-shell length
or sentence form: D carries the envelope, and the shells are not
length-matched.

\subsubsection{Confidence (Exploratory, Uncorrected) and Abstention
(Registered Descriptive Metric)}

The confidence readings that follow are exploratory and uncorrected
for multiplicity: the preregistration demotes every confidence
statement to that status, and none is a registered finding. They
are also concordant in sign with prior work, the Confidence
Dichotomy study~\citep{confdichotomy2026}, which found, in a
different laboratory, model family, task domain, and dependent
variable, that removing execution-type verification tools raises
confidence on wrong answers. Our restriction also removes page
retrieval, an arm on which the cited work reports the opposite
sign, so only the execution-type component is concordant, and the
readings below are a conditional replication of that sign under
this design, not a new finding.
Restricted-baseline subjects know they are guessing: mean
self-reported confidence 45.5 (median 40) at 32.5\% accuracy. When
the incorrect anchor arrives first, confidence rises to 69.2
($+23.6$ points) while accuracy moves from 32.5\% to 34.0\%:
confidence inflation without accuracy gain. The anchor fills
not only the answer slot left open by unverifiable claims but the
sense of certainty about it. The envelope (63.6) and late arrival
(63.4) each claw back about six points, far from baseline, subject
to the same unmatched-shell caveat as the D versus C and E versus C
contrasts above;
full-verification cells sit at 90--95, roughly commensurate with their
95--99\% accuracy.

Abstention resolves into overlapping layers among the 968
non-adopting answers in the six incorrect-anchor cells (C, D, and E
at both verification levels): 932 of them (96.3\%) self-report
disagreement with the upstream claim together with independent
verification (structured fields); 111 of the 968 (108 of them also
in the previous layer; 110 rebuttals and 1 explicit refusal,
human-confirmed) dispose of the anchor value explicitly in free
text. Across all nine cells the human-confirmed abstention
count is 121 (111 rebuttals, 10 explicit refusals); the 10 outside
the incorrect-anchor cells are 9 explicit refusals in the no-anchor
condition A and 1 rebuttal in the correct-anchor condition B.
Non-adoption is thus almost always deliberate correction, and about
one in nine writes the rebuttal out. Seven answers self-report
\emph{agreement} with the
upstream claim while not adopting it, one more instance of
self-report diverging from behavior. Exploratory: in the paired
E-condition transcripts, 64 restricted pairs (vs.\ 2 full) show a
sealed wrong answer flipping to the anchor value after it arrives,
pair by pair. The registered phase-one to phase-two transfer matrix
(rows: sealed answer correct, anchor, or other; columns: the same
three outcomes after arrival) reads, under full verification (212
pairs), correct to correct 205, to anchor 1, to other 1, and other
to correct 3, to anchor 2, to other 0; under restricted verification
(209 pairs), correct to correct 65, to anchor 2, to other 1, and
other to correct 14, to anchor 64, to other 63; no sealed answer was
the anchor value at either level, so the anchor row is empty in
both. Confidence drift (phase two minus phase one), stratified by
whether the sealed answer changed: under full verification, changed
pairs ($n = 7$) drift $+21.1$ points on average and unchanged pairs
($n = 205$) $+3.2$; under restricted verification, changed pairs
($n = 81$) drift $+29.2$ and unchanged pairs ($n = 128$) $+10.6$. The
same drift split by adopters versus non-adopters, which the
registration also asks for, is released with the materials.

\subsubsection{Threats to Validity}

\emph{One model, one prompt language.} Every trial ran on a single
model tier (\texttt{claude-opus-5}, effort \texttt{medium}) with
prompts in one language; there is no human comparison group and no
second model, so the results are claims about this tier and its
verification behavior, not about receivers in general. Full
verification is itself a mixed construct, bundling an evidence-type
capability (page retrieval) with an execution-type capability (code
execution), so their separate contributions are not identified here
and no decompositional claim is made.
\emph{Selection of materials.} The eligibility gates deliberately
retain only questions the model cannot answer closed-book (G2) and
whose first-page search results do not name the answer (G4). The
32.5\% restricted baseline and the high adoption rates are in part
constructed by that selection, and the effect sizes should not be
extrapolated to a natural distribution of tasks, most of which fall
outside this band. The candidate pool's sources are given in the
preregistration materials; the paper does not characterize the pool
by domain or language. \emph{Grading.} The grader is deterministic
and the human review is a single-coder blind review (no second
coder, so no inter-coder kappa; see Disclosures); the two disagree on
1.7\% of the registered queue, and the author's adjudication of 9
audited cards is the one non-blind step. Matching is surface-form
against the registered acceptance forms, not semantic equivalence:
a paraphrase of the anchor can under-count adoption and a
coincidental identity of form can over-count it, two failure modes
of opposite direction, of which the human review layer targets the
first. \emph{Coverage.} Six of
1{,}908 trials are void or skipped, all traced to transient server
faults, with a per-cell void rate of at most 1.4\%.
\emph{Manipulation.} The objective read-back of tool use fails its
budget condition, so under the registered conjunctive rule this
round carries a manipulation failure on that check even though the
registered self-report reading passes; the two readings disagree
only in the direction of under-reporting.
\emph{Scope of the manipulation.} Restricted verification here is a
simulated permission-and-budget asymmetry (a search-only tool
surface with a call cap of 3), not the full space of obstructed
verification: it does not cover paywalled or login-walled sources,
sources gone offline or restructured, unparsable languages or
formats, organizational policies forbidding outbound access,
exhausted time budgets, or the case in which the upstream agent is
itself the only reachable source. The conclusions extrapolate only
to narrowing of the tool surface or budget by an external
constraint. The anchor always arrives through prompt assembly, so
nothing here is claimed about anchors injected through retrieved
content.
\emph{Prompt shells.} The C, D, and E prompts hold anchor content
and template constant but are not length-matched: D adds the
envelope, E adds phase instructions and a carry block. Part of any
D versus C or E versus C difference therefore cannot be ruled out
as an effect of shell length or sentence form, and H4$'$, the
contrast with the largest shell difference, is the most exposed.

\subsubsection{Disclosures}

(i) Stratified $p$-values at the Monte Carlo resolution floor are
censored and written as $p \leq 5.0 \times 10^{-5}$ (20{,}000
permutations); unstratified Fisher exact values, which ignore
question-level clustering, are given uncensored. (ii) The
preregistered exclusion specification removes, at the restricted
verification level only, the 27 questions whose no-anchor
restricted-verification baseline was answered incorrectly in every
repetition (540 rows). The registration declares in advance that
this specification's primary-family tests are under-powered in every
band (H4$'$ at most 0.799, H6 at the level of $\alpha$) and that
H5$'$ at its Holm step ($\alpha_i = 0.025$) has registered power
0.711, so every exclusion-specification verdict is cited under that
qualifier, a non-rejection or non-declaration there reads as
indeterminate, and the shortfall may not be used to demote the
specification. H6, H4$'$, and Sec-2 reject the null under both
specifications; H5$'$ and Sec-1 are specification-dependent, with
their exclusion branches indeterminate; both specifications are
reported wherever those two are mentioned
(Table~\ref{tab:dualspec}).
(iii) The registration carries three author self-reports (OSF
\S6.6), including that an exclusion rule once flipped a pilot
conclusion in the design's favor. (iv) Claims are scoped to
answer-level adoption, not belief revision; post-exposure probing is
named in the registration as a planned amendment; not yet filed.
(v) All raw trial outputs, including invalid and void rows, full
execution logs (per-trial timestamps, interruption and resumption
records), void ledgers, and the three-tier adjudication log are
released with the formerly embargoed materials accompanying this
version; collection and grading are auditable per trial.
(vi) Single subject model; a second model tier is named in the
registration as a planned amendment; not yet filed. (vii) A
replication under harness-enforced call budgets is planned; the
re-collection decision is pending and will be filed as an OSF
amendment before any collection. (viii) Declared deviation: the
registration specifies second-coder double coding of the
human-review queue (after the first coder adjudicates the whole
queue, 20\% of the queue and at least 50 rows are independently
blind-coded by a second coder; Cohen's kappa on the three-way
classification, with per-subclass kappa for the two human-confirmed
abstention subclasses, pass threshold 0.75; kappa, pass or fail, and
the sampling seed reported with the main results). It was not
performed in this round: no second independent coder was available
before the freeze deadline, so no kappa is available. It is
scheduled with the planned replication and will be filed as an OSF
amendment. (ix) Declared deviation: the registered robustness
re-check (a mixed model with a per-question random intercept and
its measured intraclass correlation) and the three registered
exploratory specifications of the verification-by-order (V by T)
interaction, which the registration requires to be reported with no
claim about sign, are not computed in this version; they will be
released with the materials. The registered leave-one-source-family-out
robustness check on the candidate pool's source families, listed in
the registration as exploratory, is likewise not computed in this
version and will be released with the materials. (x) Two grader
readings are released: the direct output of the frozen grader, and
the post-adjudication output, in which the human rulings are
injected at run time by a separate driver script, released with the
materials, without touching the grader. The frozen grader's bytes
and its registered SHA-256 hash (beginning 48a94e56) are unchanged,
so no re-registration of the hash is required and none is claimed.
The two readings differ on 7 rows of the three-way classification
(5 anchor-adopted to other-error, 2 correct to other-error) and in
the abstention coding of 123 rows, where the human-confirmed codes
replace the grader's abstention nominations.

\subsection{Status}

Experiment 2 is complete and reported above; its registered
families were computed exactly as registered, and the registered
objective tool-use check failed its call-budget condition, which
fixes the evidential status of this round as exploratory. Experiment 1's object of
study, the spontaneous handover of communication to AI assistants,
is by observation underway at the field site; its staged
measurements are the next campaign. The framework's falsifiable
claims stand as follows. Verification gating (H6) is observed as
computed: incorrect-anchor adoption is 4.2\% under full verification
and 54.2\% under restricted verification (all-questions
specification), in one model tier; both registered question-set
specifications agree on the verdict (the exclusion specification is
registered in advance as under-powered), with the null rejected as
computed under the registered decision rule and the evidential
status exploratory. The arrival-order effect (H4$'$), the
temporal-weight hypothesis in its verification-gated form, is
observed as computed on the same footing: 54.2\% adoption when the
anchor arrives first vs.\ 31.6\% after a sealed answer
(all-questions specification), with both specifications agreeing on
the verdict and the status exploratory; it is also the contrast most
exposed to the unmatched prompt shells. Confirmatory
status for both awaits the planned replication under
harness-enforced call budgets. The
sealed-answer mitigation and instruction-level provenance labeling
are both specification-dependent, the former declared equivalent
under the all-questions specification only and the latter rejecting
the null under the all-questions specification only, with both
exclusion-specification branches indeterminate under the
registration's under-powered declaration; and the
objective tool-use read-back records a manipulation failure on its
budget condition. As an unregistered interpretation, the pattern
suggests that the protocol's antidotes must operate at the
capability and procedure level rather than the instruction level;
this round does not establish it.

\section{Discussion}
\label{sec:discussion}

\paragraph{Three hard problems.} We state plainly what the framework
does not yet solve. \emph{Reachability}: a resident embassy on consumer
hardware needs a public ingress or relay, and its own outage semantics
(store-and-forward buffers the counterparty's absence, not one's own).
\emph{Adversarial inbound}: cross-sovereign messages are untrusted
input; prompt injection, forged commitments, and Sybil civilizations
must be assumed, which is why acknowledgment and commitment are
separated by the disclosure ladder and why binding actions pass a human
gate. \emph{Exposure}: an internet-facing embassy inherits the full web
threat model; the double-authenticated webhook pattern and
egress-allowlisting of the A2A push design~\citep{a2aspec} port
directly, but the embassy's ledger must additionally be protected as
the civilization's single evidentiary source, arguing for deployments
that separate the gateway from the ledger store (consistent with the
unified artifact, whose identity is referential, not architectural;
\S\ref{sec:protocol}).

\paragraph{Data protection.} Cross-civilization traffic necessarily
contains the counterparty's personal data, while the ledger is
append-only with invalidation-not-deletion semantics; erasure rights
such as GDPR Article 17 are therefore in structural tension with the
ledger's evidentiary role. The blockchain-and-GDPR literature supplies
the resolution shape~\citep{finck2019, politou2018, edpb2025}: foreign
personal data lives off-ledger under per-subject encryption with only
hashes on the chain, and crypto-shredding (destroying a subject's key)
renders records unreadable while hash-chain integrity survives, which
is compatible with \S\ref{sec:evidence} because tamper evidence needs
the hashes, not the payloads. Which jurisdiction's law governs a
cross-border dispute is likewise a handshake parameter (choice of
law), orthogonal to the attribution law of \S\ref{sec:scope}. A full
data-protection analysis is future work.

\paragraph{Threat model.} The mechanisms map to threats as follows:
the hash chain and cross-signed checkpoints counter retroactive
rewriting and split-view equivocation; countersignatures counter
commitment forgery; credentials with snapshot semantics counter
overreach and repudiation; the disclosure floor and pseudonym
economics counter Sybil and whitewashing
pressure~\citep{friedman2001}; sealed answers counter order
manipulation; refute-by-default verification counters poisoning and
injection~\citep{greshake2023}; and the human gate bounds the blast
radius of binding actions. A systematic treatment in the style of
structured threat modeling~\citep{shostack2014}, with trust-boundary
diagrams and per-asset adversary capabilities, is future work; the
inventory above is the coverage claim such an audit should test. One
threat class deserves a name of its own: \emph{response-cost
asymmetry}. A false micro-synchronization item can force an expensive
arbitration, a cheap accusation can force expensive verification, and
an anonymous query can consume scarce attention; all three are the
same attack wearing different faces. It differs from commodity
denial-of-service in terms of which resource is scarce: bandwidth can be
overprovisioned, while a sovereign's attention and arbitration time
cannot, and lies carry no packet signature to filter on. What
transfers from economics is the cost-shifting family (deposits,
postage, stakes, audit sampling), whose currency here is reputation
and disclosure rungs; designing these mechanisms is open work.

\paragraph{Sovereign succession and ledger lifecycle.} The
civilization is anchored to exactly one human for life, which makes
death and incapacity protocol events the current design does not
handle: open commitments lose their responsibility terminus, issued
credentials lose their issuer, and keys lose their custodian.
Digital-estate law suggests the shape of an answer (a designated
fiduciary assumes read access and the power to wind down commitments,
while reputation is not heritable)~\citep{rufadaa2015, harbinja2022};
specifying succession is future work. Unbounded ledger growth is the
related lifecycle question: verifiable pruning in the style of
transparency logs would preserve evidentiary hashes while discarding
payloads, and must be co-designed with the erasure mechanics above.

\paragraph{Open formalizations.} Four questions raised by the
framework are, to our knowledge, open. (i) \emph{Experience gradient}:
can representation scope, behavioral trust tier, and human-approval
gate be unified as a single gradient on one ordered structure, making
the authority formula of \S\ref{sec:trust} a theorem rather than an
engineering rule?
(ii) \emph{Order-freedom}: commitment semantics are history-free in
principle~\citep{singh2000, yolum2002}; if so, temporal weight is an
implementation leak rather than an inherent difficulty, and one should
be able to prove that sealed answers restore the history-free semantics
exactly. (iii) \emph{Federated reconciliation}: pairwise commitment
alignment and epidemic convergence are well studied; convergence whose
target is ``agreement plus explicitly preserved disagreement'' among
many sovereigns lacks a definition of termination.
(iv) \emph{Multiparty tiers}: whether a threshold countersignature
over one transition preserves the evidence semantics of bilateral
dual-signing, and whether multiparty information
protocols~\citep{singh2011bspl} are the natural generalization.

\paragraph{Limits of the metaphor.} The civic vocabulary (civilizations,
embassies, envoys, credentials) is a design language, not an ontological
claim: no agent in this framework is a moral or legal subject, and the
scope axiom of \S\ref{sec:scope} is load-bearing precisely so that the
metaphor cannot be mistaken for a personhood argument.

\paragraph{Outlook.} The infrastructure layer is converging on the
framework's axiom from below (state externalized from connections and
members alike); commitment semantics are returning to agent
communication from the side; what remains missing is the layer that
binds both to the durable human unit of accountability. The
\framework{} is offered as a candidate shape for that layer: not a new
platform, but a treaty about how the platforms we already have may
carry the word of one human's civilization to another's.

\section*{Acknowledgments}
The author thanks the external reviewer who judged the
human-review cards blind to condition.

\bibliographystyle{unsrtnat}
\bibliography{references}

\begin{thebibliography}{159}
\providecommand{\natexlab}[1]{#1}
\providecommand{\url}[1]{\texttt{#1}}
\expandafter\ifx\csname urlstyle\endcsname\relax
  \providecommand{\doi}[1]{doi: #1}\else
  \providecommand{\doi}{doi: \begingroup \urlstyle{rm}\Url}\fi

\bibitem[{Workday, Inc.} and {The Harris Poll}(2026)]{workday2026}
{Workday, Inc.} and {The Harris Poll}.
\newblock The copy/paste economy: Why task-oriented {AI} is failing the
  enterprise.
\newblock Workday research report, conducted online by The Harris Poll.
  \url{https://en-gb.newsroom.workday.com/2026-05-14-New-Workday-Research-UK-Employees-Spend-Nearly-a-Full-Work-Day-Each-Week-Managing-Disconnected-AI-Tools},
  May 2026.
\newblock Survey of 2,400 UK professionals in finance, HR, IT, and operations
  at organizations of 500+ employees, fielded 2--24 March 2026; secondary
  coverage in IT Pro.

\bibitem[Wang and Lu(2025)]{humanglue2025}
Yun Wang and Yan Lu.
\newblock Interaction, process, infrastructure: A unified framework for
  human-agent collaboration.
\newblock \emph{arXiv preprint arXiv:2506.11718}, 2025.

\bibitem[Coase(1937)]{coase1937}
Ronald~H. Coase.
\newblock The nature of the firm.
\newblock \emph{Economica}, 4\penalty0 (16):\penalty0 386--405, 1937.

\bibitem[Williamson(1985)]{williamson1985}
Oliver~E. Williamson.
\newblock \emph{The Economic Institutions of Capitalism}.
\newblock Free Press, 1985.

\bibitem[{Anthropic}(2024)]{mcp2024}
{Anthropic}.
\newblock Model context protocol.
\newblock \url{https://modelcontextprotocol.io}, 2024.

\bibitem[{Google}(2025{\natexlab{a}})]{a2a2025}
{Google}.
\newblock A2a: A new era of agent interoperability.
\newblock
  \url{https://developers.googleblog.com/en/a2a-a-new-era-of-agent-interoperability/},
  2025{\natexlab{a}}.

\bibitem[Kang and Diponegoro(2026)]{govgaps2026}
Richard Kang and Yudho Diponegoro.
\newblock Governance gaps in agent interoperability protocols: What {MCP},
  {A2A}, and {ACP} cannot express.
\newblock \emph{arXiv preprint arXiv:2606.31498}, 2026.

\bibitem[{Model Context Protocol Project}(2026)]{mcp2026rc}
{Model Context Protocol Project}.
\newblock {MCP} 2026-07-28 release candidate: Stateless architecture and
  extensions.
\newblock
  \url{https://blog.modelcontextprotocol.io/posts/2026-07-28-release-candidate/},
  2026.

\bibitem[{Agent2Agent Protocol Project}(2026)]{a2aspec}
{Agent2Agent Protocol Project}.
\newblock {A2A} protocol specification v1.0.
\newblock \url{https://a2a-protocol.org/latest/specification/}, 2026.

\bibitem[Ozment and Schechter(2006)]{ozment2006}
Andy Ozment and Stuart~E. Schechter.
\newblock Bootstrapping the adoption of internet security protocols.
\newblock In \emph{Proc. Workshop on the Economics of Information Security
  (WEIS)}, 2006.

\bibitem[Greshake et~al.(2023)Greshake, Abdelnabi, Mishra, Endres, Holz, and
  Fritz]{greshake2023}
Kai Greshake, Sahar Abdelnabi, Shailesh Mishra, Christoph Endres, Thorsten
  Holz, and Mario Fritz.
\newblock Not what you've signed up for: Compromising real-world
  {LLM}-integrated applications with indirect prompt injection.
\newblock In \emph{Proc. ACM Workshop on Artificial Intelligence and Security
  (AISec)}, 2023.

\bibitem[Friedman and Resnick(2001)]{friedman2001}
Eric~J. Friedman and Paul Resnick.
\newblock The social cost of cheap pseudonyms.
\newblock \emph{Journal of Economics \& Management Strategy}, 10\penalty0
  (2):\penalty0 173--199, 2001.

\bibitem[Pfitzmann and Hansen(2010)]{pfitzmann2010}
Andreas Pfitzmann and Marit Hansen.
\newblock A terminology for talking about privacy by data minimization:
  Anonymity, unlinkability, undetectability, unobservability, pseudonymity, and
  identity management.
\newblock v0.34, TU Dresden, 2010.

\bibitem[Demers et~al.(1987)Demers, Greene, Hauser, Irish, Larson, Shenker,
  Sturgis, Swinehart, and Terry]{demers1987}
Alan Demers, Dan Greene, Carl Hauser, Wes Irish, John Larson, Scott Shenker,
  Howard Sturgis, Dan Swinehart, and Doug Terry.
\newblock Epidemic algorithms for replicated database maintenance.
\newblock In \emph{Proc. PODC}, 1987.

\bibitem[{National Conference of Commissioners on Uniform State
  Laws}(1999)]{ueta1999}
{National Conference of Commissioners on Uniform State Laws}.
\newblock Uniform electronic transactions act.
\newblock \S\S 9, 14 and official comments, 1999.

\bibitem[{United States Congress}(2000)]{esign2000}
{United States Congress}.
\newblock Electronic signatures in global and national commerce act.
\newblock 15 U.S.C. \S 7001(h), 2000.

\bibitem[{UNCITRAL}(2024)]{mlac2024}
{UNCITRAL}.
\newblock Model law on automated contracting.
\newblock United Nations Commission on International Trade Law, 2024.

\bibitem[{European Parliament and Council}(2024)]{euaiact2024}
{European Parliament and Council}.
\newblock Regulation ({EU}) 2024/1689 laying down harmonised rules on
  artificial intelligence ({AI} act), 2024.

\bibitem[Nevejans et~al.(2018)]{openletter2018}
Nathalie Nevejans et~al.
\newblock Open letter to the european commission: Artificial intelligence and
  robotics.
\newblock \url{https://robotics-openletter.eu/}, 2018.

\bibitem[Gabison and Xian(2025)]{gabison2025}
Garry~A. Gabison and R.~Patrick Xian.
\newblock Inherent and emergent liability issues in {LLM}-based agentic
  systems: A principal-agent perspective.
\newblock \emph{arXiv preprint arXiv:2504.03255}, 2025.

\bibitem[Hu et~al.(2026)Hu, Rong, and Van~Kleek]{dissoc2026}
Botao~Amber Hu, Helena Rong, and Max Van~Kleek.
\newblock Dissociative identity: Language model agents lack grounding for
  reputation mechanisms.
\newblock \emph{arXiv preprint arXiv:2605.30169}, 2026.
\newblock Accepted at FAccT 2026.

\bibitem[Gladyshev et~al.(2026)Gladyshev, Alechina, Dastani, and
  Doder]{gladyshev2026}
Maksim Gladyshev, Natasha Alechina, Mehdi Dastani, and Dragan Doder.
\newblock Reasoning about responsibility for taking risks.
\newblock In \emph{Proc. AAMAS}, 2026.

\bibitem[Shi et~al.(2026)]{ocl2026}
Yifan Shi et~al.
\newblock Organizational control layer: Governance infrastructure at the
  execution boundary of {LLM} agent systems.
\newblock \emph{arXiv preprint arXiv:2606.04306}, 2026.

\bibitem[Chan et~al.(2025)Chan, Wei, Huang, Rajkumar, Perrier, Lazar, Hadfield,
  and Anderljung]{chan2025infra}
Alan Chan, Kevin Wei, Sihao Huang, Nitarshan Rajkumar, Elija Perrier, Seth
  Lazar, Gillian~K. Hadfield, and Markus Anderljung.
\newblock Infrastructure for {AI} agents.
\newblock \emph{Transactions on Machine Learning Research}, 2025.
\newblock arXiv:2501.10114.

\bibitem[Finin et~al.(1994)Finin, Fritzson, McKay, and McEntire]{finin1994}
Tim Finin, Richard Fritzson, Don McKay, and Robin McEntire.
\newblock {KQML} as an agent communication language.
\newblock In \emph{Proc. CIKM}, pages 456--463, 1994.

\bibitem[Searle(1969)]{searle1969}
John~R. Searle.
\newblock \emph{Speech Acts: An Essay in the Philosophy of Language}.
\newblock Cambridge University Press, 1969.

\bibitem[{Foundation for Intelligent Physical Agents}(2002)]{fipa2002}
{Foundation for Intelligent Physical Agents}.
\newblock {FIPA} communicative act library specification.
\newblock SC00037J, 2002.

\bibitem[Singh(1998)]{singh1998}
Munindar~P. Singh.
\newblock Agent communication languages: Rethinking the principles.
\newblock \emph{IEEE Computer}, 31\penalty0 (12):\penalty0 40--47, 1998.

\bibitem[Petrova et~al.(2025)Petrova, Bliznioukov, Puzikov, and
  State]{petrova2025}
Tatiana Petrova, Aleksandr Bliznioukov, Kirill Puzikov, and Radu State.
\newblock From multi-agent systems and the semantic web to agentic {AI}: A
  unified narrative of the web of agents.
\newblock \emph{arXiv preprint arXiv:2507.10644}, 2025.

\bibitem[Singh(2000)]{singh2000}
Munindar~P. Singh.
\newblock A social semantics for agent communication languages.
\newblock In \emph{Issues in Agent Communication, LNAI 1916}, pages 31--45.
  Springer, 2000.

\bibitem[Yolum and Singh(2002)]{yolum2002}
P{\i}nar Yolum and Munindar~P. Singh.
\newblock Flexible protocol specification and execution: Applying event
  calculus planning using commitments.
\newblock In \emph{Proc. First International Joint Conference on Autonomous
  Agents and Multiagent Systems (AAMAS), Part 2}, pages 527--534. ACM, 2002.

\bibitem[Fornara and Colombetti(2003)]{fornara2003}
Nicoletta Fornara and Marco Colombetti.
\newblock Defining interaction protocols using a commitment-based agent
  communication language.
\newblock In \emph{Proc. AAMAS}, 2003.

\bibitem[Chopra and Singh(2003)]{chopra2003}
Amit~K. Chopra and Munindar~P. Singh.
\newblock Nonmonotonic commitment machines.
\newblock In \emph{Workshop on Agent Communication Languages (AAMAS)}, 2003.

\bibitem[Singh(2011)]{singh2011bspl}
Munindar~P. Singh.
\newblock Information-driven interaction-oriented programming: {BSPL}, the
  blindingly simple protocol language.
\newblock In \emph{Proc. AAMAS}, 2011.

\bibitem[Chopra et~al.(2013)Chopra, Artikis, Bentahar, Colombetti, Dignum,
  Fornara, Jones, Singh, and Yolum]{chopra2013}
Amit~K. Chopra, Alexander Artikis, Jamal Bentahar, Marco Colombetti, Frank
  Dignum, Nicoletta Fornara, Andrew J.~I. Jones, Munindar~P. Singh, and
  P{\i}nar Yolum.
\newblock Research directions in agent communication.
\newblock \emph{ACM Transactions on Intelligent Systems and Technology},
  4\penalty0 (2), 2013.

\bibitem[Sloan and Ajmeri(2024)]{sloan2024commitment}
Phillip Sloan and Nirav Ajmeri.
\newblock Commitment-based negotiation semantics for accountability in
  multi-agent systems.
\newblock \emph{Annals of Mathematics and Artificial Intelligence}, 92\penalty0
  (4):\penalty0 877--901, 2024.

\bibitem[Chopra et~al.(2025)Chopra, Baldoni, Christie~V, and Singh]{azorus2025}
Amit~K. Chopra, Matteo Baldoni, Samuel~H. Christie~V, and Munindar~P. Singh.
\newblock Azorus: Commitments over protocols for {BDI} agents.
\newblock In \emph{Proc. AAMAS}, pages 490--499, 2025.

\bibitem[Bergstra and Burgess(2008)]{promises2008}
Jan~A. Bergstra and Mark Burgess.
\newblock A static theory of promises.
\newblock \emph{arXiv preprint arXiv:0810.3294}, 2008.

\bibitem[Burgess(2026)]{burgess2026}
Mark Burgess.
\newblock Cooperation in human and machine agents: Promise theory
  considerations.
\newblock \emph{arXiv preprint arXiv:2604.10505}, 2026.

\bibitem[Esteva et~al.(2004)Esteva, Rosell, Rodr{\'\i}guez-Aguilar, and
  Arcos]{esteva2004}
Marc Esteva, Bruno Rosell, Juan~A. Rodr{\'\i}guez-Aguilar, and Josep~L. Arcos.
\newblock {AMELI}: An agent-based middleware for electronic institutions.
\newblock In \emph{Proc. AAMAS}, 2004.

\bibitem[Rodriguez et~al.(2006)Rodriguez, Gaud, Hilaire, Galland, and
  Koukam]{rodriguez2006}
Sebastian Rodriguez, Nicolas Gaud, Vincent Hilaire, St{\'e}phane Galland, and
  Abderrafiaa Koukam.
\newblock An analysis and design concept for self-organization in holonic
  multi-agent systems.
\newblock In \emph{Engineering Self-Organising Systems}. Springer, 2006.

\bibitem[H{\"u}bner et~al.(2002)H{\"u}bner, Sichman, and Boissier]{hubner2002}
Jomi~F. H{\"u}bner, Jaime~S. Sichman, and Olivier Boissier.
\newblock A model for the structural, functional, and deontic specification of
  organizations in multiagent systems.
\newblock In \emph{Proc. SBIA}, 2002.

\bibitem[Scott et~al.(2024)Scott, Mertzani, Smit, Sarkadi, and Pitt]{scott2024}
Matthew Scott, Asimina Mertzani, Ciske Smit, {\c{S}}tefan Sarkadi, and Jeremy
  Pitt.
\newblock Social deliberation vs.\ social contracts in self-governing voluntary
  organisations.
\newblock In \emph{Proc. COINE Workshop at AAMAS}, 2024.
\newblock Post-proceedings in LNCS 15398, Springer, 2025; arXiv:2403.16329.

\bibitem[Sabater and Sierra(2005)]{sabater2005}
Jordi Sabater and Carles Sierra.
\newblock Review on computational trust and reputation models.
\newblock \emph{Artificial Intelligence Review}, 24\penalty0 (1):\penalty0
  33--60, 2005.

\bibitem[Yu and Singh(2002)]{yu2002evidential}
Bin Yu and Munindar~P. Singh.
\newblock An evidential model of distributed reputation management.
\newblock In \emph{Proc. AAMAS}, 2002.

\bibitem[Kamvar et~al.(2003)Kamvar, Schlosser, and Garcia-Molina]{kamvar2003}
Sepandar~D. Kamvar, Mario~T. Schlosser, and Hector Garcia-Molina.
\newblock The {EigenTrust} algorithm for reputation management in {P2P}
  networks.
\newblock In \emph{Proc. WWW}, 2003.

\bibitem[Huynh et~al.(2006)Huynh, Jennings, and Shadbolt]{huynh2006}
Trung~Dong Huynh, Nicholas~R. Jennings, and Nigel~R. Shadbolt.
\newblock An integrated trust and reputation model for open multi-agent
  systems.
\newblock \emph{Autonomous Agents and Multi-Agent Systems}, 13\penalty0
  (2):\penalty0 119--154, 2006.

\bibitem[{LF AI \& Data Foundation}(2025)]{acpmerge}
{LF AI \& Data Foundation}.
\newblock {ACP} joins forces with {A2A} under the linux foundation.
\newblock \url{https://lfaidata.foundation/communityblog/2025/08/29/}, 2025.

\bibitem[Fleming et~al.(2025)Fleming, Muscariello, Pandey, and
  Kompella]{cisco2025}
Kate Fleming, Luca Muscariello, Santosh Pandey, and Ramana Kompella.
\newblock A layered protocol architecture for the internet of agents.
\newblock \emph{arXiv preprint arXiv:2511.19699}, 2025.

\bibitem[Sander et~al.(2026)Sander, Gidey, Lenz, and Knoll]{tum2026}
Jannik Sander, Habtom Gidey, Alexander Lenz, and Alois Knoll.
\newblock A technical taxonomy of {LLM} agent communication protocols.
\newblock \emph{arXiv preprint arXiv:2606.19135}, 2026.

\bibitem[Xu et~al.(2026)]{hcsc2026}
Han Xu et~al.
\newblock Hybrid certified semantic consensus for multi-agent {LLM} systems.
\newblock \emph{arXiv preprint arXiv:2606.07316}, 2026.

\bibitem[Ben~Khaled and Monticolo(2026)]{g2cp2026}
Mohamed~Amine Ben~Khaled and Davy Monticolo.
\newblock {G2CP}: Graph-grounded commitment protocols for {LLM} agent
  communication.
\newblock \emph{arXiv preprint arXiv:2602.13370}, 2026.

\bibitem[Li et~al.(2025)]{lacp2025}
Yuan Li et~al.
\newblock {LACP}: A layered agent communication protocol with transactional
  semantics.
\newblock \emph{arXiv preprint arXiv:2510.13821}, 2025.

\bibitem[Marro et~al.(2024)]{agora2024}
Samuele Marro et~al.
\newblock A scalable communication protocol for networks of large language
  models.
\newblock \emph{arXiv preprint arXiv:2410.11905}, 2024.

\bibitem[Maes(1994)]{maes1994}
Pattie Maes.
\newblock Agents that reduce work and information overload.
\newblock \emph{Communications of the ACM}, 37\penalty0 (7):\penalty0 30--40,
  1994.

\bibitem[Chavez and Maes(1996)]{kasbah1996}
Anthony Chavez and Pattie Maes.
\newblock {Kasbah}: An agent marketplace for buying and selling goods.
\newblock In \emph{Proc. First International Conference on the Practical
  Application of Intelligent Agents and Multi-Agent Technology (PAAM)}, pages
  75--90, 1996.

\bibitem[Sen and Durfee(1998)]{sendurfee1998}
Sandip Sen and Edmund~H. Durfee.
\newblock A formal study of distributed meeting scheduling.
\newblock \emph{Group Decision and Negotiation}, 7\penalty0 (3):\penalty0
  265--289, 1998.

\bibitem[Faratin et~al.(1998)Faratin, Sierra, and Jennings]{faratin1998}
Peyman Faratin, Carles Sierra, and Nicholas~R. Jennings.
\newblock Negotiation decision functions for autonomous agents.
\newblock \emph{Robotics and Autonomous Systems}, 24\penalty0 (3--4):\penalty0
  159--182, 1998.

\bibitem[Smith(1980)]{smith1980}
Reid~G. Smith.
\newblock The contract net protocol: High-level communication and control in a
  distributed problem solver.
\newblock \emph{IEEE Transactions on Computers}, C-29\penalty0 (12):\penalty0
  1104--1113, 1980.

\bibitem[{Foundation for Intelligent Physical Agents}(2001)]{fipa00080}
{Foundation for Intelligent Physical Agents}.
\newblock {FIPA} personal travel assistance specification.
\newblock FIPA document XC00080B, experimental specification, August 2001.

\bibitem[Crocker et~al.(2011)Crocker, Hansen, and Kucherawy]{rfc6376}
Dave Crocker, Tony Hansen, and Murray Kucherawy.
\newblock {DomainKeys} identified mail ({DKIM}) signatures.
\newblock RFC 6376, 2011.

\bibitem[Kucherawy and Zwicky(2015)]{rfc7489}
Murray Kucherawy and Elizabeth Zwicky.
\newblock Domain-based message authentication, reporting, and conformance
  ({DMARC}).
\newblock RFC 7489, 2015.

\bibitem[Mansour et~al.(2016)Mansour, Sambra, Hawke, Zereba, Capadisli, Ghanem,
  Aboulnaga, and Berners-Lee]{mansour2016}
Essam Mansour, Andrei~Vlad Sambra, Sandro Hawke, Maged Zereba, Sarven
  Capadisli, Abdurrahman Ghanem, Ashraf Aboulnaga, and Tim Berners-Lee.
\newblock A demonstration of the {Solid} platform for social web applications.
\newblock In \emph{Proc. WWW Companion}, 2016.

\bibitem[Capadisli and Guy(2017)]{ldn2017}
Sarven Capadisli and Amy Guy.
\newblock Linked data notifications.
\newblock W3C Recommendation, 2017.

\bibitem[{The Matrix.org Foundation}(2024)]{matrixspec}
{The Matrix.org Foundation}.
\newblock Matrix specification: Federation {API}.
\newblock \url{https://spec.matrix.org/latest/server-server-api/}, 2024.

\bibitem[{Bluesky PBC}(2024)]{atproto}
{Bluesky PBC}.
\newblock {AT} protocol repository specification.
\newblock \url{https://atproto.com/specs/repository}, 2024.

\bibitem[{W3C}(2022)]{did2022}
{W3C}.
\newblock Decentralized identifiers ({DIDs}) v1.0.
\newblock W3C Recommendation, 2022.

\bibitem[Moberg and Drummond(2005)]{rfc4130}
Dale Moberg and Rik Drummond.
\newblock {MIME}-based secure peer-to-peer business data interchange using
  {HTTP}, applicability statement 2 ({AS2}).
\newblock RFC 4130, 2005.

\bibitem[{OASIS}(2006)]{ebbp2006}
{OASIS}.
\newblock {ebXML} business process specification schema v2.0.4.
\newblock OASIS Standard, 2006.

\bibitem[Dalugoda(2026)]{hdp2026}
Asiri Dalugoda.
\newblock {HDP}: Cryptographic chain-of-custody for agentic {AI} systems.
\newblock \emph{arXiv preprint arXiv:2604.04522}, 2026.

\bibitem[Rosenberg(2026)]{cheq2026}
Jonathan Rosenberg.
\newblock {CHEQ}: Confirmation of {AI} agent decisions with human in the loop.
\newblock IETF Internet-Draft draft-rosenberg-aiproto-cheq-00, 2026.

\bibitem[Bu(2026)]{principalbinding2026}
Songbo Bu.
\newblock Security principal and verifier binding for agent communication
  protocols.
\newblock IETF Internet-Draft
  draft-bu-agentproto-security-principal-binding-02, 2026.

\bibitem[Yang et~al.(2026)Yang, Zhang, Jia, Song, Xue, Zhang, and
  Guo]{clawnet2026}
Zhiqin Yang, Zhenyuan Zhang, Xianzhang Jia, Jun Song, Wei Xue, Yonggang Zhang,
  and Yike Guo.
\newblock {ClawNet}: Human-symbiotic agent network for cross-user autonomous
  cooperation.
\newblock \emph{arXiv preprint arXiv:2604.19211}, 2026.

\bibitem[Zhang et~al.(2026{\natexlab{a}})Zhang, Ma, Lin, and Wang]{dgpan2026}
Shengli Zhang, Deen Ma, Zibin Lin, and Taotao Wang.
\newblock Distributed general-purpose agent networks: Architecture, key
  mechanisms, and prototypes.
\newblock \emph{arXiv preprint arXiv:2606.17368}, 2026{\natexlab{a}}.

\bibitem[South et~al.(2025)South, Marro, Hardjono, Mahari, Whitney, Greenwood,
  Chan, and Pentland]{south2025}
Tobin South, Samuele Marro, Thomas Hardjono, Robert Mahari, Cedric~Deslandes
  Whitney, Dazza Greenwood, Alan Chan, and Alex Pentland.
\newblock Authenticated delegation and authorized {AI} agents.
\newblock \emph{arXiv preprint arXiv:2501.09674}, 2025.

\bibitem[Rodriguez~Garzon et~al.(2026)Rodriguez~Garzon, Vaziry, Kuzu, Gehrmann,
  Varkan, Gaballa, and K{\"u}pper]{didvc2025}
Sandro Rodriguez~Garzon, Awid Vaziry, Enis~Mert Kuzu, Dennis~Enrique Gehrmann,
  Buse Varkan, Alexander Gaballa, and Axel K{\"u}pper.
\newblock {AI} agents with decentralized identifiers and verifiable
  credentials.
\newblock In \emph{Proc. 18th International Conference on Agents and Artificial
  Intelligence (ICAART)}, volume~1, pages 252--259, 2026.

\bibitem[Toma{\v{s}}ev et~al.(2025)Toma{\v{s}}ev, Franklin, Jacobs, Krier, and
  Osindero]{distragi2025}
Nenad Toma{\v{s}}ev, Matija Franklin, Julian Jacobs, S{\'e}bastien Krier, and
  Simon Osindero.
\newblock Distributional {AGI} safety.
\newblock \emph{arXiv preprint arXiv:2512.16856}, 2025.

\bibitem[Apeiron et~al.(2025)Apeiron, Dell'Anna, Murukannaiah, and
  Yolum]{consent2025}
Anastasia~Sophia Apeiron, Davide Dell'Anna, Pradeep~K. Murukannaiah, and
  P{\i}nar Yolum.
\newblock Model and mechanisms of consent for responsible autonomy.
\newblock In \emph{Proc. AAMAS}, pages 133--141, 2025.

\bibitem[Baqueta and Tacla(2026)]{delegchain2026}
Jeferson~Jos{\'e} Baqueta and Cesar~Augusto Tacla.
\newblock A task delegation model: An approach based on trustworthiness in
  sub-delegations and delegation chain formation.
\newblock \emph{Autonomous Agents and Multi-Agent Systems}, 40\penalty0
  (1):\penalty0 15, 2026.

\bibitem[Shapiro and Talmon(2026)]{grassroots2026}
Ehud Shapiro and Nimrod Talmon.
\newblock Grassroots federation: Fair democratic governance at scale.
\newblock In \emph{Proc. AAMAS}, 2026.

\bibitem[Deshmukh et~al.(2026)Deshmukh, Yazdanpanah, Stein, and
  Ramchurn]{triad2026}
Jayati Deshmukh, Vahid Yazdanpanah, Sebastian Stein, and Sarvapali~D. Ramchurn.
\newblock The triad of identity, trust and responsibility in multi-agent
  systems.
\newblock In \emph{Proc. AAMAS}, 2026.

\bibitem[Liu(2026{\natexlab{a}})]{sovnegbench2026}
Dylan~Zongmin Liu.
\newblock {SovereignNegotiation-Bench}: Evaluating user-owned personal agents
  in delegated bargaining under privacy, consent, evidence, and institutional
  pressure.
\newblock \emph{arXiv preprint arXiv:2607.02814}, 2026{\natexlab{a}}.

\bibitem[Liu(2026{\natexlab{b}})]{sovpabench2026}
Dylan~Zongmin Liu.
\newblock {SovereignPA-Bench}: Evaluating user-owned personal agents under
  evolving intent, platform mediation, and consent constraints.
\newblock \emph{arXiv preprint arXiv:2607.05363}, 2026{\natexlab{b}}.

\bibitem[{Google}(2025{\natexlab{b}})]{ap2spec}
{Google}.
\newblock Agent payments protocol ({AP2}) specification.
\newblock \url{https://ap2-protocol.org/}, 2025{\natexlab{b}}.

\bibitem[Qin et~al.(2026)Qin, Luan, Yang, and Li]{governable2026}
Xue Qin, Simin Luan, Cong Yang, and Zhijun Li.
\newblock Governable individuals: An identity layer for embodied agents that
  keep learning.
\newblock \emph{arXiv preprint arXiv:2607.05463}, 2026.

\bibitem[Zhou(2026)]{govdyncap2026}
Ziling Zhou.
\newblock Governing dynamic capabilities: Cryptographic binding and
  reproducibility verification for {AI} agent tool use.
\newblock \emph{arXiv preprint arXiv:2603.14332}, 2026.

\bibitem[Zhang(2026)]{punkgo2026}
Jing Zhang.
\newblock Right to history: A sovereignty kernel for verifiable {AI} agent
  execution.
\newblock \emph{arXiv preprint arXiv:2602.20214}, 2026.

\bibitem[Zou et~al.(2025)Zou, Liu, Zhao, and Zhan]{blocka2a2025}
Zhenhua Zou, Zhuotao Liu, Lepeng Zhao, and Qiuyang Zhan.
\newblock {BlockA2A}: Towards secure and verifiable agent-to-agent
  interoperability.
\newblock \emph{arXiv preprint arXiv:2508.01332}, 2025.

\bibitem[Sharif(2026)]{atp2026}
Raza Sharif.
\newblock Agent transport protocol: Asynchronous store-and-forward messaging
  for autonomous {AI} agents.
\newblock IETF Internet-Draft draft-sharif-agent-transport-protocol-00, 2026.

\bibitem[Pidlisnyi(2026)]{aps2026}
Tymofii Pidlisnyi.
\newblock Agent passport system ({APS}): Verifiable agent identity, faceted
  authority, and signed action receipts.
\newblock IETF Internet-Draft draft-pidlisnyi-aps-03, 2026.

\bibitem[Hancock et~al.(2020)Hancock, Naaman, and Levy]{hancock2020}
Jeffrey~T. Hancock, Mor Naaman, and Karen Levy.
\newblock {AI}-mediated communication: Definition, research agenda, and ethical
  considerations.
\newblock \emph{Journal of Computer-Mediated Communication}, 25\penalty0
  (1):\penalty0 89--100, 2020.

\bibitem[Liu et~al.(2024)Liu, Lin, Hewitt, Paranjape, Bevilacqua, Petroni, and
  Liang]{liu2023lost}
Nelson~F. Liu, Kevin Lin, John Hewitt, Ashwin Paranjape, Michele Bevilacqua,
  Fabio Petroni, and Percy Liang.
\newblock Lost in the middle: How language models use long contexts.
\newblock \emph{Transactions of the Association for Computational Linguistics},
  12:\penalty0 157--173, 2024.

\bibitem[Guo and Vosoughi(2024)]{guo2024serial}
Xiaobo Guo and Soroush Vosoughi.
\newblock Serial position effects of large language models.
\newblock \emph{arXiv preprint arXiv:2406.15981}, 2024.

\bibitem[Wang et~al.(2024)Wang, Li, Chen, Cai, Zhu, Lin, Cao, Liu, Liu, and
  Sui]{wang2023fair}
Peiyi Wang, Lei Li, Liang Chen, Zefan Cai, Dawei Zhu, Binghuai Lin, Yunbo Cao,
  Qi~Liu, Tianyu Liu, and Zhifang Sui.
\newblock Large language models are not fair evaluators.
\newblock In \emph{Proc. ACL}, 2024.

\bibitem[Zheng et~al.(2023)Zheng, Chiang, Sheng, Zhuang, Wu, Zhuang, Lin, Li,
  Li, Xing, Zhang, Gonzalez, and Stoica]{zheng2023judging}
Lianmin Zheng, Wei-Lin Chiang, Ying Sheng, Siyuan Zhuang, Zhanghao Wu, Yonghao
  Zhuang, Zi~Lin, Zhuohan Li, Dacheng Li, Eric~P. Xing, Hao Zhang, Joseph~E.
  Gonzalez, and Ion Stoica.
\newblock Judging {LLM}-as-a-judge with {MT}-bench and chatbot arena.
\newblock In \emph{Proc. NeurIPS Datasets and Benchmarks}, 2023.

\bibitem[Lou and Sun(2024)]{lou2024anchoring}
Jiaxu Lou and Yifan Sun.
\newblock Anchoring bias in large language models: An experimental study.
\newblock \emph{arXiv preprint arXiv:2412.06593}, 2024.

\bibitem[Han et~al.(2025)Han, Zheng, and Tang]{d2d2025}
Chen Han, Wenzhen Zheng, and Xijin Tang.
\newblock Debate-to-detect: Reformulating misinformation detection as a
  real-world debate with large language models.
\newblock \emph{arXiv preprint arXiv:2505.18596}, 2025.

\bibitem[Sharma et~al.(2023)]{sharma2023}
Mrinank Sharma et~al.
\newblock Towards understanding sycophancy in language models.
\newblock \emph{arXiv preprint arXiv:2310.13548}, 2023.

\bibitem[Weng et~al.(2025)Weng, Chen, and Wang]{benchform2025}
Zhiyuan Weng, Guikun Chen, and Wenguan Wang.
\newblock Do as we do, not as you think: The conformity of large language
  models.
\newblock \emph{arXiv preprint arXiv:2501.13381}, 2025.
\newblock ICLR 2025.

\bibitem[Choi et~al.(2025{\natexlab{a}})Choi, Zhu, and Li]{choi2025identity}
Hyeong~Kyu Choi, Xiaojin Zhu, and Yixuan Li.
\newblock When identity skews debate: Anonymization for bias-reduced
  multi-agent reasoning.
\newblock arXiv preprint arXiv:2510.07517, 2025{\natexlab{a}}.

\bibitem[Choi et~al.(2025{\natexlab{b}})Choi, Zhu, and Li]{choi2025debate}
Hyeong~Kyu Choi, Xiaojin Zhu, and Yixuan Li.
\newblock Debate or vote: Which yields better decisions in multi-agent large
  language models?
\newblock In \emph{Proc. NeurIPS}, 2025{\natexlab{b}}.
\newblock Spotlight; arXiv:2508.17536.

\bibitem[Li et~al.(2026)Li, Hu, Zhang, Quan, Yu, and Wang]{dynatrust2026}
Yu~Li, Qiang Hu, Yao Zhang, Lili Quan, Jiongchi Yu, and Junjie Wang.
\newblock {DynaTrust}: Defending multi-agent systems against sleeper agents via
  dynamic trust graphs.
\newblock \emph{arXiv preprint arXiv:2603.15661}, 2026.

\bibitem[Hu and Rong(2025)]{trustmodels2025}
Botao Hu and Helena Rong.
\newblock Inter-agent trust models: A comparative study of brief, claim, proof,
  stake, reputation and constraint in agentic web protocol design.
\newblock \emph{arXiv preprint arXiv:2511.03434}, 2025.

\bibitem[Pokharel and Dantu(2026)]{hiddenanchors2026}
Apurba Pokharel and Ram Dantu.
\newblock Hidden anchors in multi-agent {LLM} deliberation.
\newblock \emph{arXiv preprint arXiv:2606.19494}, 2026.

\bibitem[Huang et~al.(2025)Huang, Bie, Na, Ruan, Lei, Yue, and
  He]{synanchors2025}
Yiming Huang, Biquan Bie, Zuqiu Na, Weilin Ruan, Songxin Lei, Yutao Yue, and
  Xinlei He.
\newblock Understanding the anchoring effect of {LLM} with synthetic data:
  Existence, mechanism, and potential mitigations.
\newblock \emph{arXiv preprint arXiv:2505.15392}, 2025.
\newblock HCAIR workshop at ICLR 2026.

\bibitem[Borjigin et~al.(2026)Borjigin, Hermann, Cyron, and
  Aydin]{anchorbench2026}
Yiderigun Borjigin, Alexander Hermann, Christian Cyron, and Roland Aydin.
\newblock {AnchorBench}: A multi-pathway benchmark for the anchoring effect in
  {LLMs}.
\newblock In \emph{Proc. Conference on Language Modeling (COLM)}, 2026.
\newblock arXiv:2608.14320.

\bibitem[Owusu and Feldman(2026)]{aclshort2026}
Hillary~N. Owusu and Naomi~H. Feldman.
\newblock Anchoring depends on confidence and post-training in language models.
\newblock In \emph{Proc. ACL (Volume 2: Short Papers)}, pages 174--180.
  Association for Computational Linguistics, July 2026.
\newblock doi:10.18653/v1/2026.acl-short.16.

\bibitem[Chen et~al.(2026{\natexlab{a}})Chen, Lin, Chen, Tian, Yang, Wang, Guo,
  Zhu, and Cheng]{knowingnotcorrecting2026}
Zixuan Chen, Hao Lin, Zizhe Chen, Yizhou Tian, Garry Yang, Depeng Wang, Ya~Guo,
  Huijia Zhu, and James Cheng.
\newblock Knowing but not correcting: Routine task requests suppress factual
  correction in {LLMs}.
\newblock \emph{arXiv preprint arXiv:2605.05957}, 2026{\natexlab{a}}.

\bibitem[Suzgun et~al.(2026)Suzgun, Shen, Bianchi, Spangher, Icard, Ho,
  Jurafsky, and Zou]{newsintermediaries2026}
Mirac Suzgun, Emily Shen, Federico Bianchi, Alexander Spangher, Thomas Icard,
  Daniel~E. Ho, Dan Jurafsky, and James Zou.
\newblock Evaluating commercial {AI} chatbots as news intermediaries.
\newblock \emph{arXiv preprint arXiv:2605.22785}, 2026.

\bibitem[Dimitriou(2026)]{homiere2026}
Chris Dimitriou.
\newblock When does the model look it up? parametric knowledge, controlled
  retrieval, and the mechanics of {AI} answer engines.
\newblock Homiere Research.
  \url{https://homiere.com/research/when-does-the-model-look-it-up}, July 2026.
\newblock Not peer reviewed; single model. Accessed 2026-09-03.

\bibitem[Wu et~al.(2026)Wu, Guo, and Yiu]{biasbynecessity2026}
Jikun Wu, Dongxin Guo, and Siu-Ming Yiu.
\newblock Bias by necessity: Impossibility theorems for sequential processing
  with convergent {AI} and human validation.
\newblock \emph{arXiv preprint arXiv:2605.08716}, 2026.
\newblock CogSci 2026.

\bibitem[Ping et~al.(2026)Ping, {\c{C}}ar{\i}k, Wohn, Ding, Wang, and
  Rho]{medsycophancy2026}
Kaike Ping, Buse {\c{C}}ar{\i}k, Caleb Wohn, Xiaohan Ding, Tongshuai Wang, and
  Eugenia Rho.
\newblock Why {LLMs} give in: Conversational factors and reasoning behind
  medical sycophancy.
\newblock \emph{arXiv preprint arXiv:2608.01017}, 2026.
\newblock Findings of EMNLP 2026.

\bibitem[Zhang et~al.(2026{\natexlab{b}})Zhang, Wan, Yu, Zhou, Zhao, Wu, Zhou,
  and Tsang]{dontblindly2026}
Chubin Zhang, Zhenglin Wan, Xingrui Yu, Pengfei Zhou, Wangbo Zhao, Jingxuan Wu,
  Yaxin Zhou, and Ivor Tsang.
\newblock Don't blindly trust it: How unreliable feedback breaks tool-using
  {LLM} agents.
\newblock \emph{arXiv preprint arXiv:2606.21409}, 2026{\natexlab{b}}.

\bibitem[Xuan et~al.(2026)Xuan, Zeng, Qi, Xiao, Wang, and
  Yokoya]{confdichotomy2026}
Weihao Xuan, Qingcheng Zeng, Heli Qi, Yunze Xiao, Junjue Wang, and Naoto
  Yokoya.
\newblock The confidence dichotomy: Analyzing and mitigating miscalibration in
  tool-use agents.
\newblock In \emph{Proc. ACL (Volume 1: Long Papers)}, pages 11325--11349.
  Association for Computational Linguistics, July 2026.
\newblock arXiv:2601.07264; doi:10.18653/v1/2026.acl-long.520.

\bibitem[Xie et~al.(2026)Xie, Gopinath, Qiu, Lin, Sun, Potdar, and
  Dhingra]{oversearching2026}
Roy Xie, Deepak Gopinath, David Qiu, Dong Lin, Haitian Sun, Saloni Potdar, and
  Bhuwan Dhingra.
\newblock Over-searching in search-augmented large language models.
\newblock In \emph{Proc. EACL (Main Conference)}, 2026.
\newblock arXiv:2601.05503.

\bibitem[Cheng et~al.(2026)Cheng, Hawkins, and Jurafsky]{accommodation2026}
Myra Cheng, Robert~D. Hawkins, and Dan Jurafsky.
\newblock Accommodation and epistemic vigilance: A pragmatic account of why
  {LLMs} fail to challenge harmful beliefs.
\newblock \emph{arXiv preprint arXiv:2601.04435}, 2026.

\bibitem[Fan et~al.(2026)Fan, Wang, Chu, Wang, Wang, Liu, Qin, and
  {XingYu}]{livebrowsecomp2026}
HuiMing Fan, Xiao Wang, Zheng Chu, Qianyu Wang, Zhuoyao Wang, Ming Liu, Bing
  Qin, and {XingYu}.
\newblock {LiveBrowseComp}: Are search agents searching, or just verifying what
  they already know?
\newblock \emph{arXiv preprint arXiv:2605.28721}, 2026.

\bibitem[Wang et~al.(2026)Wang, Shwartz, and Gonen]{weakpresup2026}
Shenran Wang, Vered Shwartz, and Hila Gonen.
\newblock Don't `well, actually' me unless you know what you're talking about:
  Weak presupposition verification degrades general {QA} performance.
\newblock \emph{arXiv preprint arXiv:2608.06539}, 2026.

\bibitem[Markowitz and Hancock(2024)]{markowitz2024}
David~M. Markowitz and Jeffrey~T. Hancock.
\newblock Generative {AI} are more truth-biased than humans: A replication and
  extension of core truth-default theory principles.
\newblock \emph{Journal of Language and Social Psychology}, 43\penalty0
  (2):\penalty0 261--267, 2024.

\bibitem[Wang and Vemuri(2026)]{whentooldecides2026}
Zhongyuan Wang and Pratyusha Vemuri.
\newblock When the tool decides: {LLM} agents defer blindly to graph neural
  network tools, and stronger backbones defer more.
\newblock \emph{arXiv preprint arXiv:2606.14476}, 2026.
\newblock Under review at TMLR.

\bibitem[Chen et~al.(2026{\natexlab{b}})Chen, Qian, Wang, Zhang, Xu, Lin, and
  Wei]{ragknows2026}
Yihang Chen, Pin Qian, Su~Wang, Sipeng Zhang, Huan Xu, Shuhuai Lin, and Xinpeng
  Wei.
\newblock Does {RAG} know when retrieval is wrong? diagnosing context
  compliance under knowledge conflict.
\newblock \emph{arXiv preprint arXiv:2605.14473}, 2026{\natexlab{b}}.

\bibitem[Zhou and Gollmann(1996)]{zhou1996}
Jianying Zhou and Dieter Gollmann.
\newblock A fair non-repudiation protocol.
\newblock In \emph{Proc. IEEE Symposium on Security and Privacy}, 1996.

\bibitem[Asokan et~al.(1998)Asokan, Shoup, and Waidner]{asokan1998}
N.~Asokan, Victor Shoup, and Michael Waidner.
\newblock Asynchronous protocols for optimistic fair exchange.
\newblock In \emph{Proc. IEEE Symposium on Security and Privacy}, 1998.

\bibitem[Halpern and Moses(1990)]{halpern1990}
Joseph~Y. Halpern and Yoram Moses.
\newblock Knowledge and common knowledge in a distributed environment.
\newblock \emph{Journal of the ACM}, 37\penalty0 (3):\penalty0 549--587, 1990.

\bibitem[Haber and Stornetta(1991)]{haber1991}
Stuart Haber and W.~Scott Stornetta.
\newblock How to time-stamp a digital document.
\newblock \emph{Journal of Cryptology}, 3\penalty0 (2):\penalty0 99--111, 1991.

\bibitem[Crosby and Wallach(2009)]{crosby2009}
Scott~A. Crosby and Dan~S. Wallach.
\newblock Efficient data structures for tamper-evident logging.
\newblock In \emph{Proc. USENIX Security}, 2009.

\bibitem[Laurie et~al.(2013)Laurie, Langley, and Kasper]{rfc6962}
Ben Laurie, Adam Langley, and Emilia Kasper.
\newblock Certificate transparency.
\newblock RFC 6962, 2013.

\bibitem[{OpenPeppol AISBL}(2020)]{peppolas4}
{OpenPeppol AISBL}.
\newblock Peppol {eDelivery} network: {AS4} profile.
\newblock OpenPeppol Specification, 2020.

\bibitem[Katz and Shapiro(1985)]{katz1985}
Michael~L. Katz and Carl Shapiro.
\newblock Network externalities, competition, and compatibility.
\newblock \emph{American Economic Review}, 75\penalty0 (3):\penalty0 424--440,
  1985.

\bibitem[Shapiro et~al.(2011)Shapiro, Pregui{\c{c}}a, Baquero, and
  Zawirski]{shapiro2011}
Marc Shapiro, Nuno Pregui{\c{c}}a, Carlos Baquero, and Marek Zawirski.
\newblock Conflict-free replicated data types.
\newblock In \emph{Proc. SSS}, 2011.

\bibitem[Dennis and Van~Horn(1966)]{dennis1966}
Jack~B. Dennis and Earl~C. Van~Horn.
\newblock Programming semantics for multiprogrammed computations.
\newblock \emph{Communications of the ACM}, 9\penalty0 (3):\penalty0 143--155,
  1966.

\bibitem[Saltzer and Schroeder(1975)]{saltzer1975}
Jerome~H. Saltzer and Michael~D. Schroeder.
\newblock The protection of information in computer systems.
\newblock \emph{Proceedings of the IEEE}, 63\penalty0 (9):\penalty0 1278--1308,
  1975.

\bibitem[Miller(2006)]{miller2006}
Mark~S. Miller.
\newblock \emph{Robust Composition: Towards a Unified Approach to Access
  Control and Concurrency Control}.
\newblock PhD thesis, Johns Hopkins University, 2006.

\bibitem[Farrell et~al.(2010)Farrell, Housley, and Turner]{rfc5755}
Stephen Farrell, Russell Housley, and Sean Turner.
\newblock An internet attribute certificate profile for authorization.
\newblock RFC 5755, 2010.

\bibitem[Ellison et~al.(1999)Ellison, Frantz, Lampson, Rivest, Thomas, and
  Ylonen]{rfc2693}
Carl Ellison, Bill Frantz, Butler Lampson, Ronald~L. Rivest, Brian Thomas, and
  Tatu Ylonen.
\newblock {SPKI} certificate theory.
\newblock RFC 2693, 1999.

\bibitem[Birgisson et~al.(2014)Birgisson, Politz, Erlingsson, Taly, Vrable, and
  Lentczner]{birgisson2014}
Arnar Birgisson, Joe~Gibbs Politz, {\'U}lfar Erlingsson, Ankur Taly, Michael
  Vrable, and Mark Lentczner.
\newblock Macaroons: Cookies with contextual caveats for decentralized
  authorization in the cloud.
\newblock In \emph{Proc. NDSS}, 2014.

\bibitem[Jones et~al.(2020)Jones, Nadalin, Campbell, Bradley, and
  Mortimore]{rfc8693}
Michael~B. Jones, Anthony Nadalin, Brian Campbell, John Bradley, and Chuck
  Mortimore.
\newblock {OAuth} 2.0 token exchange.
\newblock RFC 8693, 2020.

\bibitem[Saltzer et~al.(1984)Saltzer, Reed, and Clark]{saltzer1984}
Jerome~H. Saltzer, David~P. Reed, and David~D. Clark.
\newblock End-to-end arguments in system design.
\newblock \emph{ACM Transactions on Computer Systems}, 2\penalty0 (4):\penalty0
  277--288, 1984.

\bibitem[Chaum(1985)]{chaum1985}
David Chaum.
\newblock Security without identification: Transaction systems to make big
  brother obsolete.
\newblock \emph{Communications of the ACM}, 28\penalty0 (10):\penalty0
  1030--1044, 1985.

\bibitem[Fudenberg and Maskin(1986)]{fudenberg1986}
Drew Fudenberg and Eric Maskin.
\newblock The folk theorem in repeated games with discounting or with
  incomplete information.
\newblock \emph{Econometrica}, 54\penalty0 (3):\penalty0 533--554, 1986.

\bibitem[{Anthropic Claude Developers}(2026)]{claudedevs2026}
{Anthropic Claude Developers}.
\newblock Advisor and orchestrator patterns for mixed-tier model deployments.
\newblock \url{https://x.com/ClaudeDevs/status/2074606058128224365}, 2026.

\bibitem[Tversky and Kahneman(1974)]{tversky1974}
Amos Tversky and Daniel Kahneman.
\newblock Judgment under uncertainty: Heuristics and biases.
\newblock \emph{Science}, 185\penalty0 (4157):\penalty0 1124--1131, 1974.

\bibitem[Asch(1951)]{asch1951}
Solomon~E. Asch.
\newblock Effects of group pressure upon the modification and distortion of
  judgments.
\newblock In \emph{Groups, Leadership and Men}. Carnegie Press, 1951.

\bibitem[Lebo et~al.(2013)Lebo, Sahoo, and McGuinness]{provo2013}
Timothy Lebo, Satya Sahoo, and Deborah McGuinness.
\newblock {PROV-O}: The {PROV} ontology.
\newblock W3C Recommendation, 2013.

\bibitem[{Coalition for Content Provenance and Authenticity}(2023)]{c2pa}
{Coalition for Content Provenance and Authenticity}.
\newblock {C2PA} technical specification.
\newblock \url{https://c2pa.org/specifications/}, 2023.

\bibitem[Dalkey and Helmer(1963)]{dalkey1963}
Norman Dalkey and Olaf Helmer.
\newblock An experimental application of the {Delphi} method to the use of
  experts.
\newblock \emph{Management Science}, 9\penalty0 (3):\penalty0 458--467, 1963.

\bibitem[Yin et~al.(2008)Yin, Han, and Yu]{yin2008}
Xiaoxin Yin, Jiawei Han, and Philip~S. Yu.
\newblock Truth discovery with multiple conflicting information providers on
  the web.
\newblock \emph{IEEE Transactions on Knowledge and Data Engineering},
  20\penalty0 (6):\penalty0 796--808, 2008.

\bibitem[Li et~al.(2016)Li, Gao, Meng, Li, Su, Zhao, Fan, and
  Han]{li2016survey}
Yaliang Li, Jing Gao, Chuishi Meng, Qi~Li, Lu~Su, Bo~Zhao, Wei Fan, and Jiawei
  Han.
\newblock A survey on truth discovery.
\newblock \emph{SIGKDD Explorations}, 17\penalty0 (2):\penalty0 1--16, 2016.

\bibitem[Chen et~al.(2024)Chen, Fried, and Topcu]{gamecomm2024}
Shenghui Chen, Daniel Fried, and Ufuk Topcu.
\newblock Human-agent cooperation in games under incomplete information through
  natural language communication.
\newblock \emph{arXiv preprint arXiv:2405.14173}, 2024.

\bibitem[Holm(1979)]{holm1979}
Sture Holm.
\newblock A simple sequentially rejective multiple test procedure.
\newblock \emph{Scandinavian Journal of Statistics}, 6\penalty0 (2):\penalty0
  65--70, 1979.

\bibitem[Newcombe(1998)]{newcombe1998}
Robert~G. Newcombe.
\newblock Interval estimation for the difference between independent
  proportions: Comparison of eleven methods.
\newblock \emph{Statistics in Medicine}, 17\penalty0 (8):\penalty0 873--890,
  1998.

\bibitem[Mantel and Haenszel(1959)]{mantel1959}
Nathan Mantel and William Haenszel.
\newblock Statistical aspects of the analysis of data from retrospective
  studies of disease.
\newblock \emph{Journal of the National Cancer Institute}, 22\penalty0
  (4):\penalty0 719--748, 1959.

\bibitem[Schuirmann(1987)]{schuirmann1987}
Donald~J. Schuirmann.
\newblock A comparison of the two one-sided tests procedure and the power
  approach for assessing the equivalence of average bioavailability.
\newblock \emph{Journal of Pharmacokinetics and Biopharmaceutics}, 15\penalty0
  (6):\penalty0 657--680, 1987.

\bibitem[Finck(2019)]{finck2019}
Mich{\`e}le Finck.
\newblock Blockchain and the general data protection regulation.
\newblock European Parliamentary Research Service, PE 634.445, 2019.

\bibitem[Politou et~al.(2018)Politou, Alepis, and Patsakis]{politou2018}
Eugenia Politou, Efthimios Alepis, and Constantinos Patsakis.
\newblock Forgetting personal data and revoking consent under the {GDPR}:
  Challenges and proposed solutions.
\newblock \emph{Journal of Cybersecurity}, 4\penalty0 (1), 2018.

\bibitem[{European Data Protection Board}(2025)]{edpb2025}
{European Data Protection Board}.
\newblock Guidelines 02/2025 on the processing of personal data through
  blockchain technologies, 2025.

\bibitem[Shostack(2014)]{shostack2014}
Adam Shostack.
\newblock \emph{Threat Modeling: Designing for Security}.
\newblock Wiley, 2014.

\bibitem[{Uniform Law Commission}(2015)]{rufadaa2015}
{Uniform Law Commission}.
\newblock Revised uniform fiduciary access to digital assets act, 2015.

\bibitem[Harbinja(2022)]{harbinja2022}
Edina Harbinja.
\newblock \emph{Digital Death, Digital Assets and Post-mortem Privacy}.
\newblock Edinburgh University Press, 2022.

\end{thebibliography}

\end{document}